\documentclass[11pt,a4paper]{article}

\usepackage{amsfonts}
\usepackage{multirow}
\usepackage{color}
\usepackage{graphicx}
\usepackage{float}
\usepackage{amsmath}
\usepackage{amssymb}
\usepackage[colorlinks=true,linkcolor=blue,citecolor=blue]{hyperref}%
\usepackage{subcaption}
\usepackage{caption}
\usepackage{comment}
\usepackage{adjustbox} 
\usepackage{tikz}
\usetikzlibrary{shapes.geometric, arrows}
\usepackage[numbers, sort&compress]{natbib}

\title{\textbf{BPS and semi-BPS kink families in two-component scalar field theories with fourth-degree polynomial potentials}}

\author{A. Alonso-Izquierdo$^{1,2}$, M. A. González León$^{1,2}$, A. González-Parra$^{2}$, J. Martín-Vaquero$^{1,2}$\\
\\
    $^1$ Departamento de Matemática Aplicada, University of Salamanca, Casas del Parque 2, 37008 -\\
Salamanca, Spain.\\
$^2$ IUFFyM, University of Salamanca, Plaza de la Merced 1, 37008 - Salamanca, Spain
}

\date{}

\begin{document}

\maketitle

\begin{abstract}
We perform a systematic study of kink solutions in two-component scalar field theories in $(1+1)$ dimensions with interaction terms of at most quartic order. Our approach is based on the Bogomolny formalism, constructing scalar potentials from suitable superpotentials and analyzing the corresponding first-order equations. While cubic polynomial superpotentials naturally generate quartic interactions, we show that more general functional forms also lead to admissible models within the same class. In this way, we identify new models supporting continuous families of kinks with nontrivial internal structure, such that they can be interpreted as composite configurations formed by multiple localized energy lumps.
\end{abstract}

\section{Introduction}

Topological defects are solutions arising in non-linear scalar field theories whose energy density is localized, so they behave as extended particles. In general, this kind of configurations cannot decay into the vacuum of the theory because of the existence of a conserved charge, the topological charge, accounting for the topological structure of the configuration space. Moreover, their non-perturbative insights are of current interest in a wide range of areas; for instance, explaining the existence of walls and branes in Condensed Matter \cite{Abrikosov1957, ConMat1,ConMat2}, Cosmology \cite{cosm}, Biochemistry \cite{BQ}, Optics \cite{OPT}, etc. 

In this paper we shall focus on kinks, topological defects in $(1+1)$-dimensional scalar field theories. Prototypical models such as the $\phi^4$ theory already exhibit a remarkably rich dynamical behavior despite involving only a single real scalar field, largely due to their non-integrable character \cite{KSC1,KSC2,KSC3,KSC4}. The extension to theories with multiple scalar components considerably enlarges the landscape of possible static solutions, and the systematic search for kink configurations in such models has been an active and fruitful area of research for several decades \cite{FTNM11,Bazeia1995:pla,KVariety}. In this broader setting, obtaining explicit analytical expressions for the defects is of particular relevance, as it provides deeper insight into their structure, stability, and interactions. Several methodological frameworks have been proven to be effective in addressing this problem. One prominent approach is based on Hamilton–Jacobi separability of the associated mechanical analogue, which ensures complete Liouville integrability and allows for the explicit construction of kink orbits \cite{KVariety,Alonso2000,KM3C}. Alternative techniques, such as deformation procedures, have also generated extensive families of analytically tractable models and solutions \cite{DEFT1,DEFT2,DEFT3,DEFT4,DEFT5,DEFT6}. Another powerful and widely used method is the Bogomolny arrangement \cite{BOGO}, which reduces the determination of static kink solutions to solving a system of first-order ordinary differential equations. This construction relies on the existence of a superpotential function whose gradient (in the internal field space) determines the scalar potential as one half of the squared norm of that gradient. In single-field models, such a superpotential can always be constructed and is uniquely defined up to an additive constant. In contrast, for multi-component theories its existence is not guaranteed. Indeed, depending on the structure of the model, there may be no superpotential at all, or alternatively one, two, or even several non-trivially related superpotentials \cite{KVariety,Ito1985,WS}. 

From a theoretical perspective, scalar field theories involving several components and polynomial interactions of at most quartic order occupy a distinguished position in both classical and quantum frameworks. In quantum field theory, polynomial potentials up to fourth degree lead to perturbatively renormalizable models in $(1+3)$ dimensions. This requirement strongly constrains the admissible self-interactions and singles out quartic terms as the highest-order operators compatible with power-counting renormalizability. Consequently, multi-field $\phi^4$-type theories provide a natural arena in which nontrivial interactions, symmetry-breaking patterns, and topological sectors can be consistently explored without introducing uncontrollable ultraviolet divergences. Beyond renormalizability, quartic polynomial potentials arise naturally as effective descriptions near critical points in statistical and condensed-matter systems. Within the Landau–Ginzburg paradigm, the free energy is expanded as a polynomial in the order parameters, and truncation at fourth order is typically sufficient to capture spontaneous symmetry breaking, phase transitions, and the structure of degenerate vacua. In models with several scalar fields, this framework accommodates rich vacuum manifolds and symmetry-breaking schemes, leading to a wide variety of topological defects and interface solutions. The quartic truncation thus represents the minimal setting in which multiple isolated minima can be obtained while retaining analytical tractability.

In the present work, we undertake a systematic analysis of two-component scalar field theories in $(1+1)$-dimensional spacetime with interaction terms of at most quartic order. Our strategy is to construct the models through an appropriate choice of superpotential functions, from which the scalar potentials follow via the standard first-order (Bogomolny) framework. While cubic polynomial superpotentials naturally generate quartic scalar interactions, they do not exhaust all possible realizations of this class of models. Indeed, more intricate functional forms can also lead to potentials of degree no higher than four, and part of our aim is to clarify and organize these different possibilities within a unified setting. In order to avoid an excessive proliferation of parameters, we impose a discrete $\mathbb{Z}_2 \times \mathbb{Z}_2$ symmetry, corresponding to reflections of each scalar field component. This assumption significantly constrains the admissible interactions while preserving a sufficiently rich phenomenology. It is worth emphasizing that translations or rotations in the internal field space allow for straightforward generalizations of the models considered here, without altering the structure of the solutions. Particular attention will be devoted to those models that admit one-parameter families of kink solutions. Such families necessarily exhibit two zero modes in the spectrum of small fluctuations: one associated with translational invariance in physical space, and a second one reflecting the existence of a continuous degeneracy within the family of static solutions itself. Our interest in this subclass of theories is motivated by the especially rich structure of their topological defects. In these cases, kinks develop an internal structure and can be interpreted as composite objects formed by several localized energy lumps, whose relative arrangement is controlled by the continuous parameter labeling the family. This internal moduli structure endows the defects with additional dynamical features and makes them particularly appealing from both mathematical and physical perspectives.

Naturally, the framework outlined above encompasses models that have already been studied in the literature, mainly the models commonly referred to as MSTB and BNRT. However, the central objective of the present work is to determine whether the known examples exhaust all possible two-component scalar field theories in $(1+1)$ dimensions with at most quartic interactions and $\mathbb{Z}_2 \times \mathbb{Z}_2$ symmetry that admit families of kink solutions. Our analysis shows that this is not the case. The constructive procedure based on the superpotential approach enables us to identify new models that have not been previously considered and that, nevertheless, support continuous families of topological defects. Although our primary emphasis will be placed on these novel systems, we will also revisit the classical models already reported in the literature, with the aim of highlighting and contrasting the distinct internal structures that kink families may develop when the scalar potential incorporates different types of interactions. To emphasize the relevance of the previously mentioned classic models some historical aspects are included:

\begin{enumerate}
\item The model now known as the MSTB system was originally introduced by Montonen in 1976 in the context of complex scalar field theories with global $U(1)$ symmetry, during his search for charged solitons \cite{Montonen1976}. In the parameter range $\sigma\in(0,1)$, two distinct types of topological kinks were identified: one-component solutions, in which one field vanishes and the remaining component reproduces the standard $\phi^4$ kink profile, and genuinely two-component solutions constrained to elliptical orbits in internal space. Related solitary waves had already appeared in earlier work by Rajaraman and Weinberg \cite{Rajaraman1975}, situating the model at the crossroads of emerging studies on multi-field defects. Shortly thereafter, a detailed stability analysis by Sarker, Trullinger and Bishop clarified the structure of the topological sector \cite{Trullinger1976}. Although the embedded $\phi^4$ kink remains a static solution of the two-field theory, it becomes unstable in the enlarged configuration space, whereas the two-component kinks are energetically stable. Further analyses based on small fluctuations were carried out in \cite{Currie1979}. In parallel, Rajaraman discovered a non-topological kink solution for $\sigma=\tfrac{1}{2}$ \cite{Rajaraman1979}, later shown to belong to a one-parameter family existing for $\sigma\in(0,1)$ \cite{Subbaswamy1980,Subbaswamy1981}. These solutions were proven to be unstable and to satisfy an energy sum rule relating their total energy to that of the topological kinks. A decisive step in the mathematical understanding of the MSTB system was achieved when Magyari and Thomas demonstrated the complete integrability of the associated mechanical analogue \cite{Magyari1984}. The static field equations were shown to be Hamilton–Jacobi separable in elliptic coordinates, enabling Ito to obtain implicit expressions for the full variety of kink orbits and to establish instability properties via Morse index arguments \cite{Ito1985,Ito1985b}. This geometrical viewpoint was further developed in a series of works by Mateos-Guilarte, where the kink manifold was interpreted in terms of geodesics of the Maupertuis–Jacobi action and analyzed through Morse theory \cite{Guilarte1987,Guilarte1988,Guilarte1992}. Subsequent research revealed that the MSTB model is not an isolated example but rather a prototype within a wider class of multi-component scalar theories, which was generalized in different scenarios \cite{Alonso1998, Alonso2000,Alonso2002c, Alonso2008}. Scattering processes involving different kinks, their excitations and decay channels in this model have been addressed in \cite{Alonso2019, Alonso2023PhysD}. The semiclassical quantization of the stable topological kinks has been considered in \cite{Alonso2002b, JARAH1, JARAH2}.

\item The BNRT models has been discussed by Bazeia and coworkers in the references \cite{Bazeia1995:pla, Bazeia1997:jopa}, where the authors identify a pair of topological kinks. Shifman and Voloshin showed that this family of systems can be found as the dimensional reduction of a generalized Wess-Zumino model with two chiral super-fields \cite{Shifman1998b:prd,Shifman1998:prd}. They found that the static kink manifold in this model comprises a one-parameter family of energy degenerate composite kinks. These solutions are formed by two basic kinks which belong to different topological sectors and whose centers are placed at distinct points. An explicit demonstration of the stability of some of these solutions is presented in \cite{Dias2007:ijompa}. In addition to this, Sakai and Sugisaka analytically explore the existence of bound states of wall-antiwall pairs \cite{Sakai2002:prd}. A supersymmetric version of this model compatible with local supersymmetry, where the kinks of the (1+1)-dimensional model promote to exact extended solutions of ${\cal N}=1$ (3+1)-dimensional supergravity was constructed in \cite{Eto2003:prd}. In this framework, the coupling of this scalar field theory model to gravity in (4+1)-dimensions in warped spacetime is considered by Bazeia in \cite{Bazeia2004:johep}. The formation of planar networks of topological defects is addressed in \cite{Bazeia2000:prd,Avelino2009:prd}. The breaking of the classical energy degeneracy for the static kink family by quantum-induced interactions has been studied in \cite{Alonso2004:npb, Alonso2014:jhep}. A first study of the kink dynamics for this model was accomplished in \cite{Alonso2002:prd} in the adiabatic approximation, where it is assumed that the kink motion is very slow while a more general framework is considered in \cite{Alonso2018:PhysD}.
\end{enumerate}
The historical overview of the aforementioned models clearly illustrates both the sustained interest they have generated and their broad range of applications. From integrability and stability analyses to supersymmetric extensions, gravitational couplings, quantum corrections, and dynamical studies, these systems have served as a fertile testing ground for a wide variety of analytical and physical ideas. At the same time, their rich moduli structure and the diversity of kink families they support highlight the structural possibilities inherent in multi-component scalar field theories with quartic interactions. This perspective naturally motivates the search for new models within the same general framework, capable of incorporating different interaction patterns and giving rise to novel configurations in the space of kink solutions. The present work is situated precisely in this context.

The paper is organized as follows. In Section 2 we present the general framework of multi-component scalar field theories in $(1+1)$ dimensions, together with the working hypotheses adopted throughout the article. In particular, we introduce the notions of BPS and semi-BPS kinks and establish the basic tools that will be used in the subsequent analysis. Section 3 is devoted to a systematic search for models satisfying our symmetry and structural restrictions, focusing first on those generated by polynomial superpotentials of degree three. In Section 4 we extend this construction to models associated with irrational superpotentials possessing a single singular point. Within this class, we identify new theories and perform a detailed study of their kink manifolds and stability properties. Section 5 develops an analogous analysis for models derived from irrational superpotentials with two singularities. In Section 6 we investigate the phenomenon that we refer to as confluence, namely the possibility that a given scalar potential can be generated by more than one superpotential. This feature implies that a single model may accommodate several distinct kink families and, consequently, may exhibit stationary solutions with a significantly richer internal structure. Finally, Section 7 summarizes our main results and outlines several directions for future research.

\section{BPS and semi-BPS kinks in two-component scalar field theories} 

Let us consider a two-component field theory in a $(1+1)$-Minkowskian space-time. The dynamics of this model is governed by the action
\begin{equation}
    S[\phi_1,\phi_2]=\int\,d^2x\Big(\sum_{a=1}^2\frac{1}{2}\partial_{\mu}\phi_a\partial^{\mu}\phi_a-U(\phi_1,\phi_2)\Big),
    \label{eq:Action}
\end{equation}
where $\phi_a\in\text{Maps}(\mathbb{R},\mathbb{R})$, $a\in\{1,2\}$, are real scalar fields and Einstein's summation criterion is applied for space-time indices. The Minkowski metric is chosen to be $\eta_{\mu\nu}=\text{diag}\{1,-1\}$ and the notation $x^0\equiv t,\,x^1\equiv x$ will be used from now on. Furthermore, the potential term $U(\phi_1,\phi_2)$ is considered to be a positive semi-definite function.

The Euler-Lagrange equations for the action functional \eqref{eq:Action} read
\begin{equation}
\frac{\partial^2\phi_1}{\partial\,t^2}-\frac{\partial^2\phi_1}{\partial\,x^2}=-\frac{\partial U}{\partial\,\phi_1} \hspace{0.5cm},\hspace{0.5cm}
\frac{\partial^2\phi_2}{\partial\,t^2}-\frac{\partial^2\phi_2}{\partial\,x^2}=-\frac{\partial U}{\partial\,\phi_2}
\label{eq:FieldEQ}
\end{equation}
and the conserved energy for a particular field configuration is
\begin{equation}
    E[\phi_1,\phi_2]=\int_{-\infty}^{\infty}\,dx\Big[\frac{1}{2} \sum_{a=1}^2\Big(\frac{\partial \phi_a}{\partial\,t}\Big)^2 + \frac{1}{2} \sum_{a=1}^2\Big(\frac{\partial \phi_a}{\partial\,x}\Big)^2 + U(\phi_1,\phi_2)\Big]\equiv \int_{-\infty}^{\infty}\,dx\,\mathcal{E}[\phi_1(x^{\mu}),\phi_2(x^{\mu})].
\end{equation}
Here, $\mathcal{E}[\phi_1(x^{\mu}),\phi_2(x^{\mu})]$ is the energy density carried by the field. The configuration space is defined as the set of finite energy maps from space-time to the internal space, i.e. $\mathcal{C}=\{\Phi(x^{\mu})\equiv (\phi_1(x^{\mu}),\phi_2(x^{\mu}))\,:\,E[\Phi(x^{\mu})]\,<\,\infty\}$. Necessary conditions for a configuration to belong to the configuration space are
\begin{equation}
    \lim_{x\to\pm\infty}\frac{\partial\Phi(x^{\mu})}{\partial\,t}=\lim_{x\to\pm\infty}\frac{\partial\Phi(x^{\mu})}{\partial\,x} = 0,\qquad \text{and}\qquad \lim_{x\to\pm\infty}\Phi(x^{\mu})\in\mathcal{M}, \label{eq:Asymptotic}
\end{equation}
where $\mathcal{M}$ is the vacuum manifold of the model constituted by the set of degenerate zeros of the potential, 
\begin{equation}
    \mathcal{M}= \left\{v_i=(v^1_i,v^2_i)\in\mathbb{R}^2\,:\,U(v_i)=0,\,\,i\in 1,...,n \right\}
\end{equation}
with $n$ being the total number of vacua. Here, we are considering $\mathcal{M}$ as a discrete set.\newline
Every element in $\mathcal{M}$ is a constant solution of \eqref{eq:FieldEQ} with zero energy. Moreover, the asymptotic value of the field is a constant of motion since any variation would require an infinite amount of energy. Therefore, the configuration space is the union of $2^n$ disconnected topological sectors, $\mathcal{C}=\sqcup_{i,j}^n\,\mathcal{C}_{ij}$, where the subindex $i,j$ refers to the vacuum reached at $x\to -\infty$ and $x\to +\infty$ respectively.

We are interested in solutions travelling at constant speed, i.e solitary waves. For this purpose, the invariance of \eqref{eq:Action} under Lorentz transformations allows us to search for static configurations $(\phi_1(x),\phi_2(x))$, and hence, the field equations are reduced to the system of ODEs
\begin{equation}
    \frac{d^2\phi_1}{d\,x^2}=\frac{\partial U}{\partial\,\phi_1},\qquad   \frac{d^2\phi_2}{d\,x^2}=\frac{\partial U}{\partial\,\phi_2}.\label{eq:E-L-S}
\end{equation}
Re-interpreting $x$ as a time coordinate and considering $(\phi_1,\phi_2)$ as the coordinates of a particle moving on $\mathbb{R}^2$, the system \eqref{eq:E-L-S} is nothing but the classical equations of motion of a particle under the potential $V = -U$, and the static energy functional now plays the role of the mechanical action,
\begin{equation}
    E[\phi_1,\phi_2]=\int_{-\infty}^{\infty}\,dx\Big[\frac{1}{2} \sum_{a=1}^2\Big(\frac{\partial \phi_a}{\partial\,x}\Big)^2 +  U(\phi_1,\phi_2)\Big] \,\, .
    \label{eq:SEF}
\end{equation}
In the context of models involving two scalar fields, the Bogomolny procedure can be applied whenever the potential $U(\phi_1,\phi_2)$ can be expressed in terms of a superpotential $W(\phi_1,\phi_2)$ in the form 
\begin{equation}
    U(\phi_1,\phi_2)=\frac{1}{2}\Big(\frac{\partial W}{\partial\,\phi_1}\Big)^2+\frac{1}{2}\Big(\frac{\partial W}{\partial\,\phi_2}\Big)^2. \label{eq:potential02}
\end{equation}
This relation ensures, by construction, that the potential $U(\phi_1,\phi_2)$ is non-negative and that the vacua of the system correspond to the critical points of the superpotential $W(\phi_1,\phi_2)$. If the superpotential is differentiable, the static energy functional \eqref{eq:SEF} can be rewritten as
\begin{align}
    & E[\phi_1,\phi_2]=\int_{-\infty}^{\infty}\,dx\sum_{a=1}^2\Big(\frac{d \phi_a}{d\,x} + (-1)^\alpha \frac{\partial W}{\partial\,\phi_a}\Big)^2 + T  \hspace{0.4cm} ,
\end{align}
where $\alpha=0,1$ and $T$ is a topological quantity defined as
\[
T = \left| \int_{v_i}^{v_j} dW \right|
\]
The minimal energy solutions are those that saturate the BPS bound, i.e. those satisfying the system of first order ODEs
\begin{equation}
    \frac{d\phi_1}{dx}=(-1)^\alpha \frac{\partial W}{\partial \,\phi_1},\qquad \frac{d\phi_2}{dx}=(-1)^\alpha \frac{\partial W}{\partial \,\phi_2};
    \label{eq:BOGO}
\end{equation}
and their energies depend solely on the topological sector to which they belong,
\begin{equation}
    E[\phi_1,\phi_2]= T = | W(\Phi(\infty))-W(\Phi(-\infty)) | \hspace{0.3cm}.
    \label{eq:KinkEN}
\end{equation}
The solutions that satisfy the first order system \eqref{eq:BOGO} connecting two different vacua are named as \textit{BPS kinks}. The factor $(-1)^\alpha$ appearing in \eqref{eq:BOGO} can be interpreted in two distinct ways: it can be absorbed into the spatial coordinate $x$, in which case kink- and antikink-type solutions become connected through a spatial reflection by the same equations. Alternatively, this factor may be incorporated directly into the superpotential $W$, leading to two distinct superpotentials, namely $W$ and $\widetilde{W}=-W$. 

In certain models the superpotential is not differentiable at one or more points; nevertheless, the Bogomolny construction can be applied piecewise. In such cases, if $\Phi_0=\Phi(x_0)$ is a singular point through which the solution passes, the energy must be expressed in the form 
\begin{align}
    & E[\phi_1,\phi_2]=\int_{-\infty}^{x_0}\,dx\sum_{a=1}^2\Big(\frac{d \phi_a}{d\,x} + (-1)^\alpha \frac{\partial W}{\partial\,\phi_a}\Big)^2 + \int_{x_0}^{\infty}\,dx\sum_{a=1}^2\Big(\frac{d \phi_a}{d\,x} + (-1)^\beta \frac{\partial \widetilde{W}}{\partial\,\phi_a}\Big)^2 + T  \hspace{0.4cm} , \label{enegia2}
\end{align}
where $W$ and $\widetilde{W}$ can denote any of the sign-related versions of the superpotential. Each segment of the solution then complies with a different first-order differential equation, and the total energy depends not only on the vacua but also on the singular points of the superpotential,
\begin{equation}
E[\phi_1,\phi_2]= T = | W(\Phi(\infty))-W(\Phi_0) | + | W(\Phi(-\infty))-W(\Phi_0) |\hspace{0.3cm}.\label{eq:SEMIENERGY}
\end{equation}
In these cases, we will refer to them as \textit{semi-BPS solutions}.

In any case, the BPS or semi-BPS kink trajectories on the $\phi_1$-$\phi_2$ plane comply with the first order differential equation
\begin{equation}
    \frac{d\phi_1}{d\phi_2}=\frac{\partial W/\partial \phi_1}{\partial W/\partial \phi_2}.
    \label{eq:ORBEQ}
\end{equation}
for one or more superpotentials. 

In this article, we focus on identifying models that follow the scheme outlined above and give rise to families of kinks of both BPS and semi-BPS type. To enable comparison across the different models, we require that the projection of their potentials onto the $\phi_1$-axis reproduces the standard $\phi^4$-model, that is, 
\begin{equation}
    U(\phi_1,\phi_2)=\frac{1}{2}(1-\phi_1^2)^2+\phi_2^2\,f(\phi_1,\phi_2);
    \label{eq:POTG}
\end{equation}
where $f(\phi_1,\phi_2)$ is a positive semi-definite second degree polynomial. For the sake of simplicity we will also require a $\mathbb{Z}_2\times\mathbb{Z}_2$ symmetry in the internal space, that is, the potential is invariant under the reflection transformations: $U((-1)^a \phi_1, (-1)^b \phi_2) = U(\phi_1,\phi_2)$ for $a,b=0,1$. Consequently, all models considered here have at least two vacua located at 
\begin{equation}
A_{\pm} = (\pm 1,0) \label{eq:vacua}
\end{equation}
and they always admit the usual $\tanh(x)$-kink connecting these vacua, that is, the expression 
\begin{equation}
    \mathcal{K}_1^{(\alpha)}(x)=\left((-1)^{\alpha} \tanh \bar{x}, \, 0  \right)\, \label{TK1}
\end{equation}
which corresponds to a one non-null component kink solution of the model. Here, $\alpha= 0$ refers to the kink solution and $\alpha=1$, to the antikink. Moreover, $\bar{x}\equiv x-x_0$, $x_0$ is an integration constant that determines the center of the kink. The subindex indicates that the solution contains just one non-null component within the axis $\phi_1$ on the internal plane $\phi_1$-$\phi_2$. From now on, we will refer to the both kink or antikink solutions with one non-null component in the $i$ axis as $\mathcal{K}_{i}(x)$, whereas to those connecting the vacua $v_i, v_j$ as $\mathcal{K}_{v_iv_j}(x)$.

The central question addressed in this work is how this one-component scalar field theory (the $\phi^4$ model) can be generalised to give rise to kink families when an additional internal degree of freedom $\phi_2$ is introduced. Notice that there is no loss of generality in considering potentials of the form \eqref{eq:POTG}, since any other polynomial expression with at least two vacua is connected via rotations and translations in the $\phi_1$-$\phi_2$ plane with the expression \eqref{eq:POTG}. 

Following the previous framework, we propose two classes of superpotentials. Since the potentials we aim to investigate are quartic polynomials, one natural approach is to consider cubic superpotentials and classify the resulting models under the required constraints. An alternative route (leading to families of semi-BPS kinks) is to introduce superpotentials that contain irrational functions with singular points. This second construction yields entirely novel models that, to the best of our knowledge, have not been previously explored in the literature.

\section{Models with polynomial superpotentials}

In this section, we impose a polynomial ansatz for the superpotential
\begin{equation}
    W(\phi_1,\phi_2)=\frac{1}{3}\phi_1^3+a_1\,\phi_1+a_2 \,\phi_1^2\,\phi_2+ a_3 \,\phi_1\,\phi_2^2 + a_4 \,\phi_2^3+a_5 \,\phi_2,\qquad a_i \in\mathbb{R},
    \label{eq:SUPPOL}
\end{equation}
where the coefficient of the leading term has been fixed in order to reproduce the structure of expression \eqref{eq:POTG}, namely the coefficient of the quadratic term of the $\phi^4$-potential. By substituting the form \eqref{eq:SUPPOL} into relation \eqref{eq:potential02} and enforcing the previously introduced $\mathbb{Z}_2\times \mathbb{Z}_2$-symmetry constraints, several of the coefficients $a_i$ become fixed or can be expressed in terms of the remaining free parameters. This procedure allows us to identify different one-parameter families of superpotentials and their corresponding potentials, which we list below:

\begin{itemize}
    \item \textbf{Case 1:} \textit{Wess-Zumino type model}: In this case the one-parameter family of superpotentials 
\begin{equation}
W_1(\phi_1,\phi_2)=\frac{\alpha \, \phi _1 \left(-\phi _1^2+3 \phi _2^2+3\right)+\phi _2 \left(-3 \phi _1^2+\phi _2^2+3\right)}{3 \sqrt{\alpha ^2+1}}, \hspace{0.5cm} \alpha\in \mathbb{R}, \label{eq:SUPP1}
\end{equation}
leads to a single potential of the form
\begin{equation}
U_1(\phi_1,\phi_2)=\frac{1}{2} \left(1-\phi _1^2\right)^2+\frac{1}{2} \left(1+\phi _2^2\right)^2+\phi _1^2 \phi _2^2-\frac{1}{2};\label{eq:POTP1}
\end{equation}
Therefore, although the potential $U_1(\phi_1,\phi_2)$ is generated by a superpotential containing the free parameter $\alpha$, the resulting potential is a polynomial expression with purely numerical coefficients. In other words, $U_1(\phi_1,\phi_2)$ corresponds to a single potential that nevertheless admits infinitely many superpotentials. This phenomenon is well known in the context of Wess–Zumino models \cite{WS1} and is associated with the superpotential being harmonic. Indeed, if the internal space $\mathbb{R}^2$ is interpreted as a complex plane via the standard identifications $\phi_1 = {\rm Re}(z)$ and $\phi_2={\rm Im}(z)$, the superpotential becomes holomorphic 
\[
W(z)=\frac{1}{3\sqrt{\alpha^2+1}}(z^3-3\,z),
\]
see \cite{WS2,WS}. Wess-Zumino models have been thoroughly studied since they constitute the first interacting quantum field theory with linearly realized supersymmetry and because of their relation to low-energy effective field theories of supersymmetric QCD, \cite{WS3}. In this (1+3) dimensional context, the model admits domain-wall solutions, which are nothing but static three-dimensional embeddings of one-dimensional kinks, \cite{WS4,WS5,WS6,WS7,WS8}. It is well known that the orbits of the kink solutions in this case are given by the expression $e^{i\theta} {\rm Re}\,W(z) = C$, see \cite{WS}, where the phase $\theta$ and the integration constant $C$ must be chosen such that the orbit passes through the vacuum points. The potential (\ref{eq:POTP1}) has only the two previously established vacua $A_\pm=(\pm 1,0)$ and the kink variety reduces in this case to the $\phi^4$-kink (\ref{TK1}). 

Additionally, it has been found that the potential $U_1(\phi_1,\phi_2)$ is a member of the well-known family of MSTB potentials $U_1(\phi_1,\phi_2) = \frac{1}{2} (\phi_1^2+ \phi_2^2-1)^2 +\frac{1}{2} \sigma^2 \phi_2^2$ with $\sigma=2$, that is, $U_1(\phi_1,\phi_2) = \frac{1}{2} (\phi_1^2+ \phi_2^2-1)^2 + 2 \phi_2^2$, see \cite{Montonen1976, Trullinger1976}. 

\item \textbf{Case 2:} \textit{BNRT type models:} In this case, the superpotential arises in the form
\begin{equation}
   W_2(\phi_1,\phi_2)=\frac{1}{3} \phi _1 \left(3 \,\beta \, \phi _2^2+\phi _1^2-3\right)\,,\qquad \beta\in\mathbb{R},\label{eq:SUPP2} 
\end{equation}
which defines the one-parametric family of potentials
\begin{equation}    
    U_2(\phi_1,\phi_2)=\frac{1}{2}(1-\phi_1^2)^2+\frac{1}{2}(1-\beta\,\phi_2^2)^2+\beta\,(1+2\beta)\,\phi_1^2\,\phi_2^2-\frac{1}{2}.\label{eq:POTP2}
\end{equation}
This potential is equivalent to the one that was mentioned before, the BNRT model, which was  proposed in \cite{Bazeia1997:jopa}. For example, the expression in \cite{Bazeia1997:jopa} is simply reproduced by rescaling $\phi_1\to\,\phi_1/2$, $\phi_2\to\,\phi_2/2$, $x^{\mu}\to 2\sqrt{2}\,x^{\mu}$ and identifying the parameters $\sigma = 2\,\beta$. The second order differential equations for the static solutions read 
\begin{align}
    & \frac{d^2\phi_1}{dx^2} = 2 \phi_1 \, \left[ \beta \, (2\, \beta +1)\, \phi_2^2+ \phi_1^2-1\right] \, ,\\
    & \frac{d^2\phi_1}{dx^2} = 2\, \beta\,  \phi_2 \left[ (2\, \beta +1)\, \phi_1^2+\beta\, \phi_2^2-1\right] \,.\nonumber
\end{align}
The solutions of this family of models have been exhaustively studied in the literature by means of the gradient equations
\begin{equation}
    \frac{d\phi_1}{dx} =\beta \, \phi _2^2+\phi _1^2-1 \hspace{0.5cm},\hspace{0.5cm} \frac{d\phi_2}{dx} = 2\, \beta  \phi _1 \phi _2 \hspace{0.2cm}, \label{eq:BPSequation2}
\end{equation}
derived from the saturation of the Bogomolny bound associated to the superpotential (\ref{eq:SUPP2}), see \cite{KVariety, Bazeia1997:jopa, Alonso1998, Guilarte1987, Guilarte1988, Guilarte1992}. We can identify two regimes in this model depending on the values of the coupling constant $\beta$. 
\begin{itemize}
    \item If $\beta<0$ the points $A_\pm = (\pm 1,0)$ constitute the only vacua of the model. In this case it can be proved that again the kink variety restricts to the solution \eqref{TK1}. Consequently, this case is of no interest within the present context.

   \item However, if $\beta>0$ two new vacua located at points 
    \[
    B_\pm = \left( 0, \pm \frac{1}{\sqrt{\beta}} \right)
    \]
   can be found in addition to the predefined vacua $A_\pm$. Consequently, the model contains a total of four vacua ${\cal M}=\{ A_-,A_+,B_-,B_+\}$. The integration of the orbit equations derived from \eqref{eq:BPSequation2} depends on the coupling constant $\beta$, leading to the expressions
\begin{equation}
    \begin{array}{ccc} (2 \beta +1) \,\kappa \, |\phi _2|^{\frac{1}{\beta }}-\beta\,  \phi _2^2+(2 \beta -1)\,   \phi _1^2-2\, \beta +1=0 &\mbox{ if } & \beta\neq 1/2 , \\[0.2cm]
    \phi _1^2-\phi _2^2 \left(\kappa +\log|\phi _2|\right)-1=0 &\mbox{ if } & \beta = 1/2,
   \end{array} \label{eq:BNRTorb}
  \end{equation}
where $\kappa\in\mathbb{R}$ is the integration constant. Each member of the one-parameter family of curves \eqref{eq:BNRTorb} starts at one of the vacuum points $A_+$ or $A_-$. However, not all values of the integration constant $\kappa$ are valid for these curves to reach the opposite vacuum $A_-$ or $A_+$ and form a kink-type solution. If the constant $\kappa$ exceeds the critical integration constant $\kappa_c$
\begin{equation}
    \kappa_c(\beta)=\begin{cases}
        & -\frac{2 \beta ^{\frac{1}{2 \beta }+1}}{2 \beta
            -1}\qquad \qquad\,\,\,\,\,\,\beta\neq 1/2,\\[0.2cm]
        & -\frac{1}{2} (1+\log 2)\qquad\,\,\,\,\,\beta=1/2.
    \end{cases}
\end{equation}
the solutions departing from a vacuum point goes to infinity in the internal plane. Conversely, all curves in \eqref{eq:BNRTorb} with $\kappa < \kappa_c$ constitute kinks in the $AA$ topological sector, that is, they asymptotically connect the vacua $A_+$ and $A_-$. In the particular case where $\kappa= \kappa_c$, the solutions live in a different topological sector, the $AB$ sector, which connects either $A_+$ or $A_-$ with the vacua $B_+$ or $B_-$. While the former comprise a (non-trivial) one-parameter family of kink solutions, the latter consist of eight solutions (counting kinks and antikinks). The orbits of these solutions are depicted in Figure \ref{fig:BNRTAA}, where the families of solutions in the $AA$ sector are shown in blue for various values of $\kappa$, and the solutions in the $AB$ sector are represented in red. 
\begin{figure}[h]
    \centering
    \includegraphics[width=6.5cm]{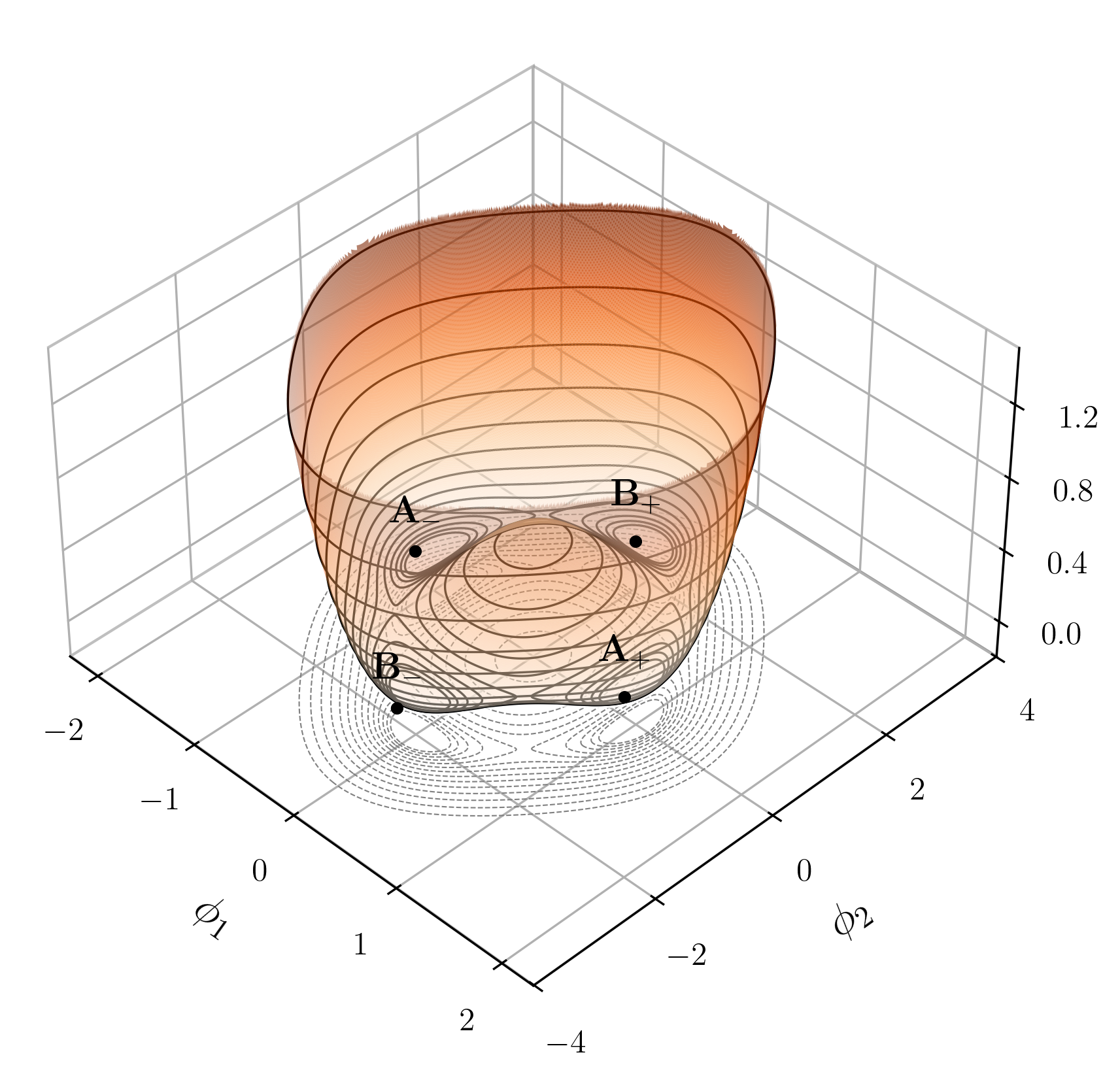}
    \includegraphics[width=6.5cm]{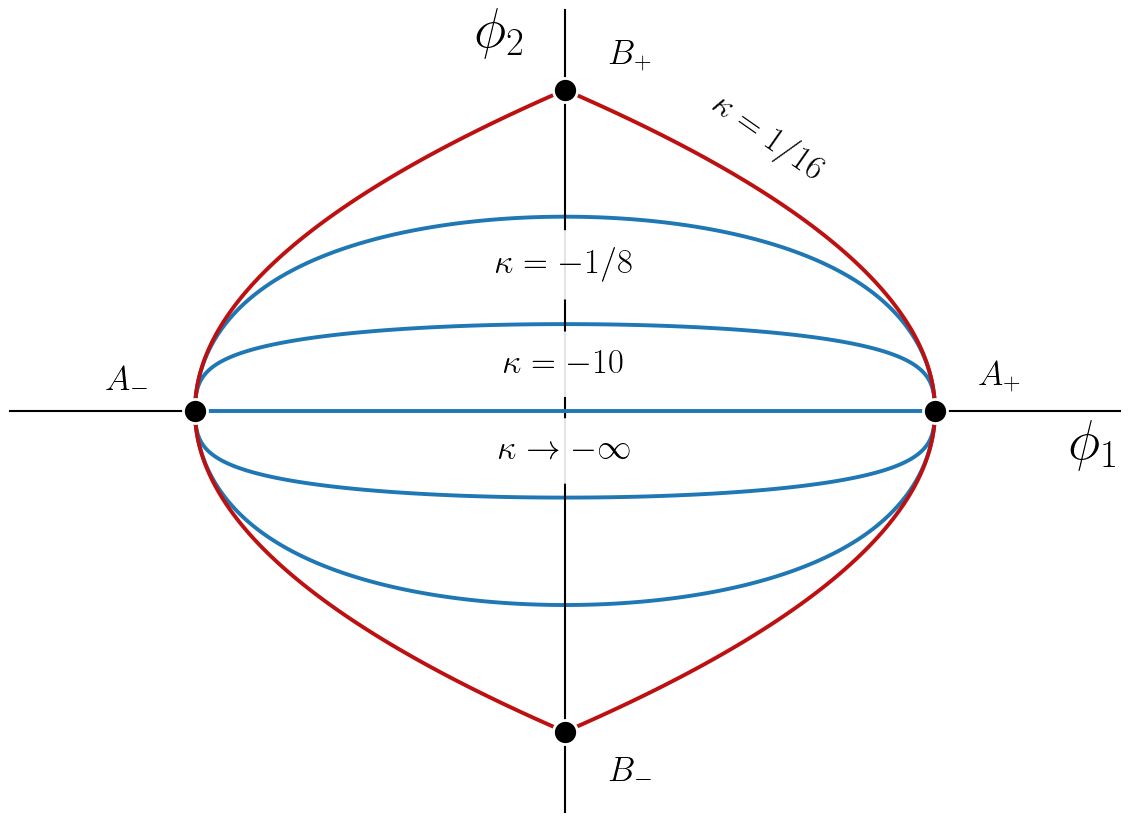}
    \caption{Potential $U_2$ (left) and its kink orbits \eqref{eq:BNRTorb} (right) for the coupling constant $\beta = 1/4$ and several values of the family parameter $\kappa$. Blue curves correspond to kinks in the topological sector $AA$ while the curves in red determine kinks in the sector $AB$ characterized by the value $\kappa=\kappa_c= 1/16$. The black dots represent the vacua of the model.}
    \label{fig:BNRTAA}
\end{figure}

From the orbit equation, \eqref{eq:BNRTorb}, we could solve for one of the fields, say $\phi_2=g(\phi_1)$, so that we are able to substitute this expression into the first order problem \eqref{eq:BPSequation2} and obtain $\phi_1(x)$. Nonetheless, the solution for the resulting first order ODE cannot be found analytically in general. As an exception, if we take $\kappa = 0$ in \eqref{eq:BNRTorb}, the orbit becomes an ellipse,
\begin{equation}
    \frac{\beta\,\phi_2^2}{1-2\,\beta} + \phi_1^2=1;
    \label{eq:BNRTellip}
\end{equation}
and explicit expressions for the profiles are known,
\begin{equation}
    \mathcal{K}_{A\overline{A}}^{(\alpha,\beta)}(x)=\left((-1)^{\alpha}\,\tanh2\,\beta\,\bar{x},(-1)^{\beta}\sqrt{\frac{2\,\beta - 1}{\beta}}\,\text{sech}\,2\,\beta\,\bar{x}\right).
\end{equation}
Here, $\alpha,\beta=0,1$, and the subindex $A\overline{A}$ means that the kink connects one of the vacua $A_{\pm}$ with the other one $A_{\mp}$ given place to a topological solution joining different vacua. However, notice that this solution is only valid within the range $\beta\in(0,1/2)$, because for $\beta>1/2$ the orbit \eqref{eq:BNRTellip} becomes an hyperbola. Some $AA$ kink profiles are depicted in Figure \ref{fig:BNRTAAprof} along with their energy densities. Notice that, depending on the value of $\kappa$, there are in general two regions at which the energy density is concentrated. Each lump can be interpreted as an extended particle and, consequently, the $AA$ topological kinks can be interpreted as composite solutions with $\kappa$ being a parameter related to the separation between basic particles. This is confirmed by computing the energy of the solutions,
\begin{equation}
    E[\mathcal{K}_{A\overline{A}}(x;\kappa)]=2E[\mathcal{K}_{AB}(x)]=\frac{4}{3}.
\end{equation}
Hence, the solutions in the $AA$ sector are composed by two kinks in the $AB$ sector.

\begin{figure}[]
  \centering
    \includegraphics[width=5cm]{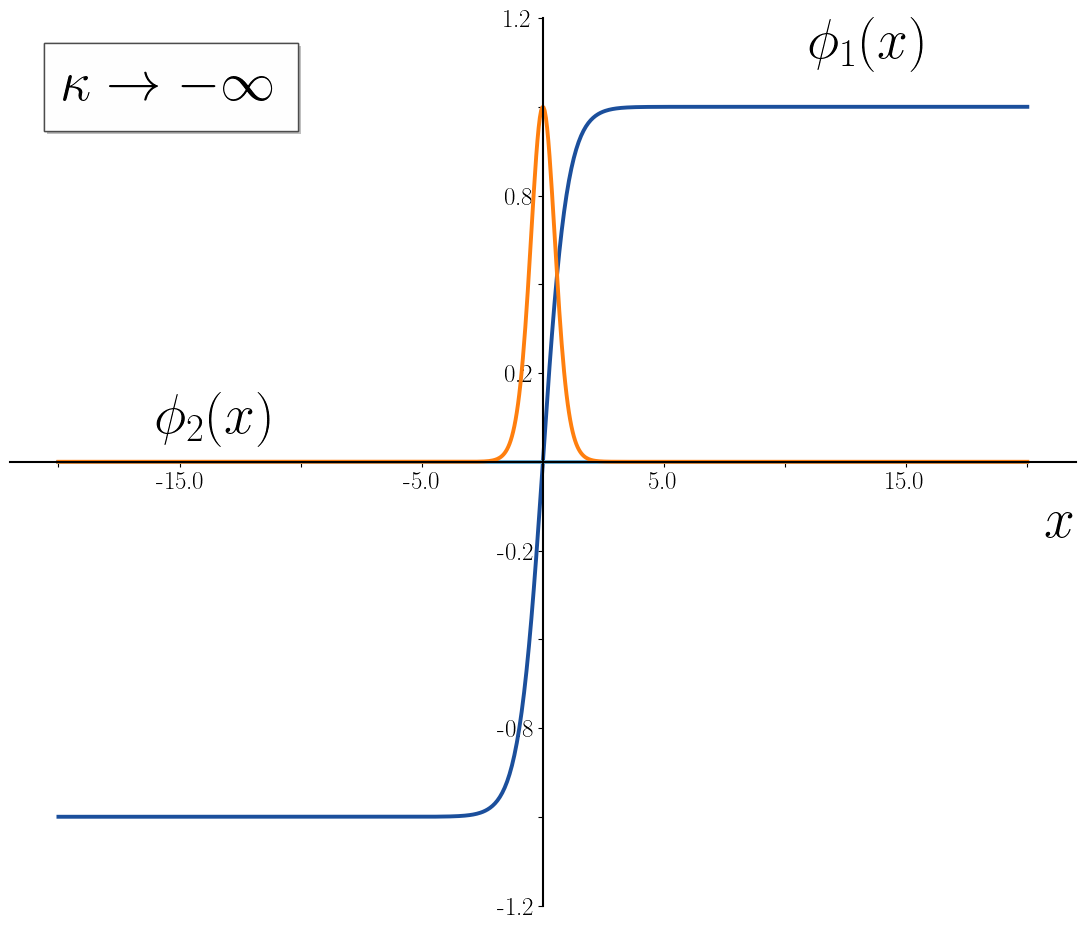} 
    \includegraphics[width=5cm]{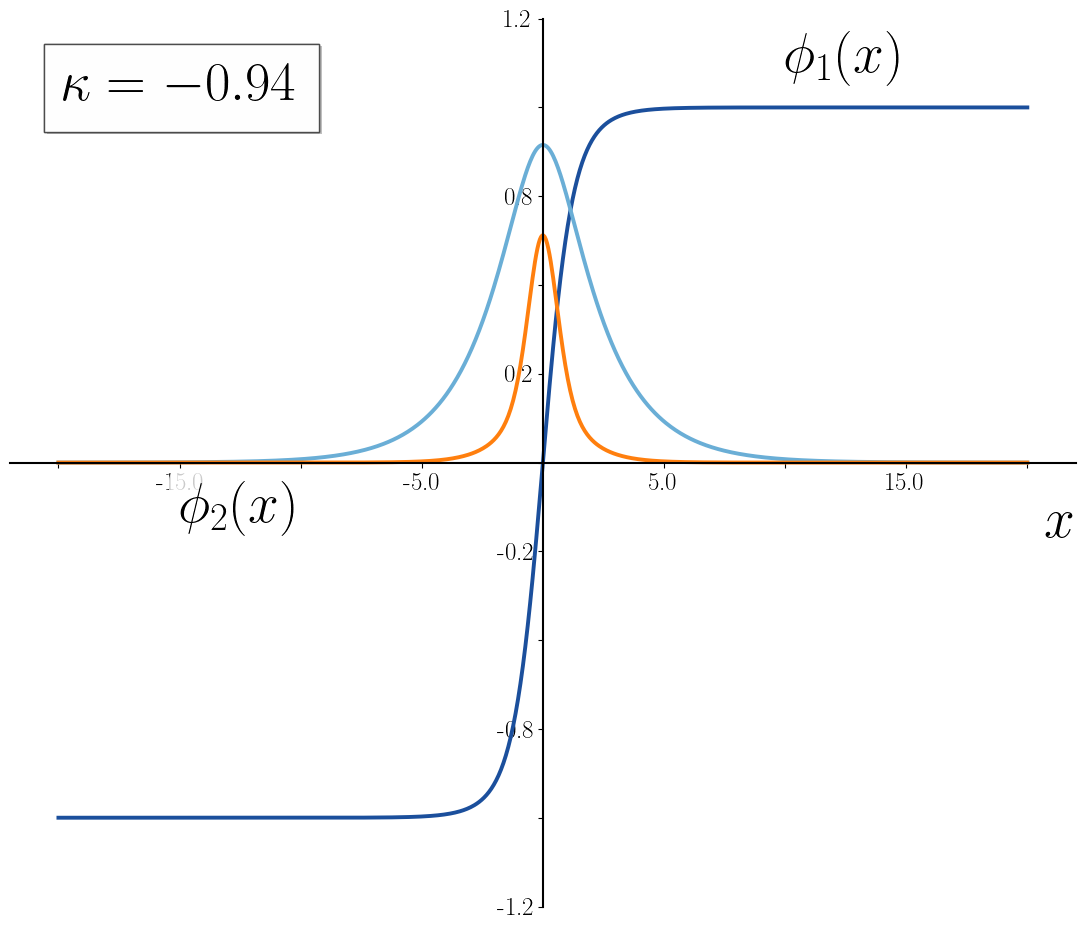}
    \includegraphics[width=5cm]{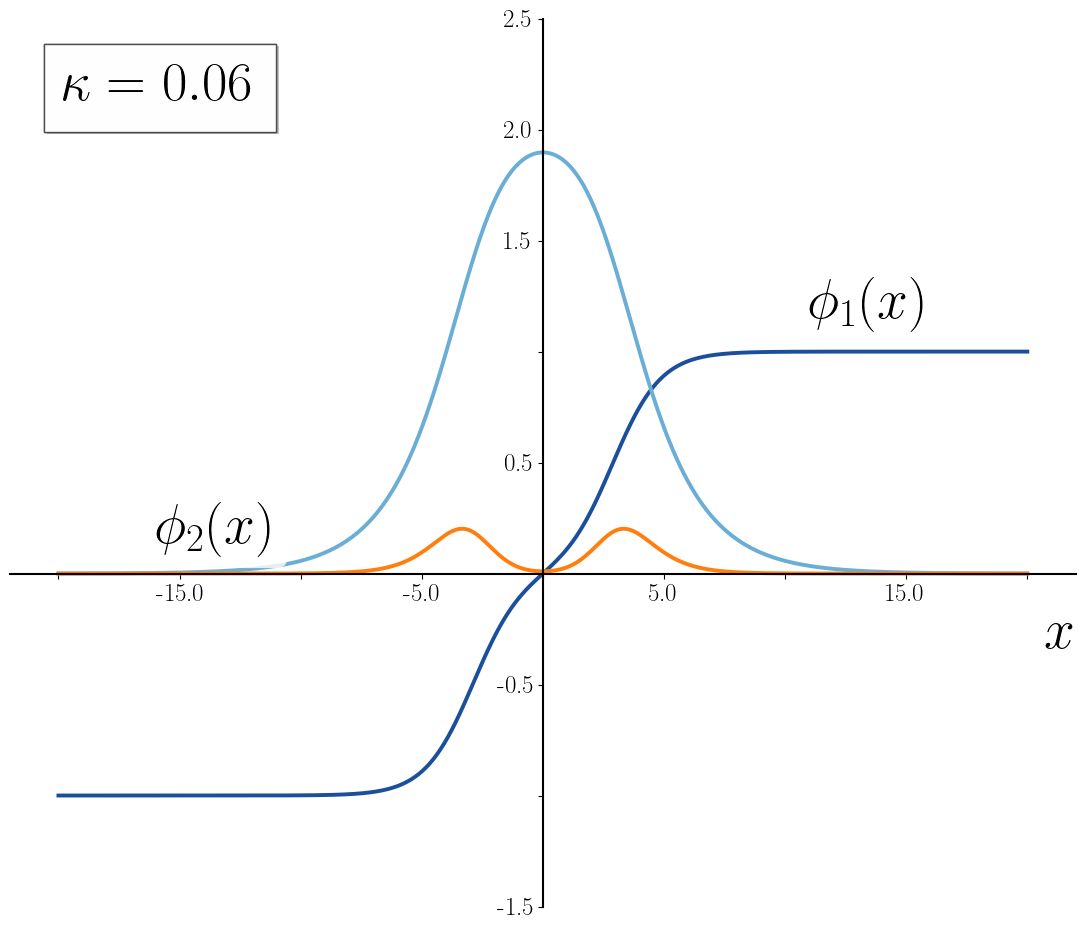}
  \caption{Graphical representation of the profiles $\Phi(x)$ for the orbits $\kappa\to-\infty$ ($\mathcal{K}_1(x)$) (left), $\kappa=-0.94$ (center) and $\kappa = 0.06$ (right) at $\beta = 1/4$. Here, $\phi_1(x)$ is the dark blue curve and $\phi_2(x)$ is the light blue one. The orange curve is the energy density associated with each profile.}
  \label{fig:BNRTAAprof}
\end{figure}

The composite nature of the $AA$ solutions makes it interesting to study their linear stability. For this purpose, we will take advantage of the fact that the kink $\mathcal{K}_1(x)$ belongs to this family for every value of $\beta>0$ to analyze its stability. In this case, the spectral problem
\begin{equation}
        \mathcal{H}_{ij}[\Phi(x)]=-\delta_{ij}\frac{d^2}{dx^2}+\frac{\partial^2U_2}{\partial\phi_i\partial\phi_j}[\Phi(x)],\qquad i,j\in\{1,2\}.
        \label{eq:Hess}
    \end{equation}
is constituted by two decoupled Pöschl-Teller potentials, making it solvable analytically;
\begin{equation}
    \mathcal{H}[\mathcal{K}_1(x)]=\begin{pmatrix}
        -\frac{d^2}{dx^2} + 4 -6\,\text{sech}^2\bar{x} & 0\\
        0                                              & -\frac{d^2}{dx^2} + 4\, \beta^2\, -2 \beta(1+2 \beta)\, \text{sech}^2\bar{x}
    \end{pmatrix}\,.
\end{equation}
The longitudinal modes are formed by the fluctuations around the usual $\phi^4$ kink: a zero mode, $F_0(x)=(\text{sech}^2x,0)$, accounting for the translational invariance of the kink, an excited shape mode, $F_1(x)=(\text{sech}\,x\,\tanh x,0)$ with eigenvalue $\omega_1^2= 3$ and a continuum spectrum with $\omega^2_1=4+k_1^2$, where $k_1\in\mathbb{R}$.

On the other hand, the transversal fluctuations spectrum contains a discrete set of eigenvalues
\begin{equation}
    \overline{\omega}_n^2=n\,(4\,\beta - n)\,,\qquad n=0,1,...,\text{E}[2\,\beta],
\end{equation}
where $\text{E}[r]$ means the integer part of $r$. The first eigenvalue is zero and, therefore, the spectrum of linear perturbations around the $\mathcal{K}_1(x)$ always comprises two zero modes. This phenomenon occurs because the $AA$ kinks constitute a bi-parametric family of solutions. Hence, shifting either the center of the kink, longitudinal perturbation, or the family parameter $\kappa$, transversal perturbation; maps solutions into solutions. As a matter of fact, we could identify when a topological kink belongs to a family of solutions by analyzing the appearance of two zero modes in its spectrum of linear fluctuations. The next shape mode are located at $\overline{\omega}_1^2=4\,\beta - 1$ for $\beta\geq 1/4$, $\overline{\omega}_2^2 = 8\beta-4$ for $\beta \geq 1/2$, etc. Thus, we can conclude that the $\mathcal{K}_1(x)$-solution is stable for every $\beta>0$.

The kink solutions described above belong either to the $AA$ topological sector (where the kinks asymptotically connect the points $A_+$ and $A_-$ or to the $AB$ sector (joining a vacuum point $A_\pm$ with another in $B_\pm$). A question that immediately arises is whether solutions exist in the remaining sector, namely the $BB$ sector, in which the vacua $B_+$ and $B_-$ are asymptotically connected. The first-order equations derived from the superpotential $W_2$ do not provide information regarding this possibility. However, a single kink solution in this sector can be readily identified from the second-order equations; specifically, it corresponds to the solution localized along the $\phi_2$-axis, whose expression is given by:
\begin{equation}
    \mathcal{K}_2^{(\alpha)}(x) = \left(0,\frac{(-1)^{\alpha}}{\sqrt{\beta}}\tanh \,\sqrt{\beta}\,\bar{x}\right)
\end{equation}
The linear stability analysis for this particular solution leads to the identification of the eigenvalues of the matrix differential operator
\begin{equation}
    \mathcal{H}[\mathcal{K}_2(x)]=\left(
\begin{array}{cc}
 -\frac{d^2}{dx^2} + 4\beta -2 (2\beta+1) \text{sech}^2\left(\sqrt{\beta }\, \bar{x}\right) & 0 \\
 0 & -\frac{d^2}{dx^2} + 4\, \beta -6\, \beta\,  \text{sech}^2\left(\sqrt{\beta }\, \bar{x}\right) \\
\end{array}
\right)\,.
\end{equation}
The longitudinal spectrum is identical to the studied in the $\mathcal{K}_1(x)$ case. However, the transversal spectrum is composed of a discrete set of eigenvalues
\begin{equation}
    \overline{\omega}_n^2 = 4\beta-\frac{\beta}{4} \left(\sqrt{\frac{8}{\beta }+17}-2 n-1\right)^2\,,\qquad \,n=0,1,...,\text{E}\left[\frac{1}{2} \left(\sqrt{\frac{8}{\beta }+17}-1\right)\right]\,.
\end{equation}
For this case, the lowest eigenvalue $\overline{\omega}_{0}^2$ is always negative, so the solution $\mathcal{K}_2(x)$ is unstable. Additionally, there exists a second zero mode for the particular values 
\begin{equation}
    \beta = \frac{2}{n^2+5n+2}\,,\qquad n\in\mathbb{N}\,, \label{casesBNRT}
\end{equation}
and, therefore, we would expect to find a second (non-trivial) one-parametric kink family in this model for the particular values
\[
\beta = 1,\, \frac{1}{4},\, \frac{1}{8},\, \frac{1}{13}, \dots  
\]
presumably derived from first order differential equations coming from a different superpotential than $W_2$.

\end{itemize}

\item Furthermore, from the aforementioned conditions in Section 2, the form of the following superpotential 
\begin{equation}
    \overline{W}(\phi_1,\phi_2)=\frac{1}{3} \phi _2 \left(-\gamma \, \phi _2^2-3 \phi _1^2+3\right) \label{eq:SUPP3} 
\end{equation}
is also obtained as a solution, which yields the potential written as follows:
\begin{equation}
   \overline{U}(\phi_1,\phi_2) = \frac{1}{2}(1-\phi_1^2)^2+\frac{1}{2}(1-\gamma\,\phi_2^2)^2+(\gamma+2)\,\phi_1^2\,\phi_2^2-\frac{1}{2},\label{eq:POTP3} 
\end{equation}
In this case, however, the symmetry involving the interchange of the fields $\phi_1$ and $\phi_2$, combined with a suitable rescaling, leads us back to the Case 2 discussed above. Specifically, for $\beta>0$ the transformation $\phi_1\leftrightarrow \phi_2$ together with the redefinition $\beta=1/\gamma$ recovers the expression in \eqref{eq:POTP2}. This fact is associated with the possibility, given by the general equations, of obtaining the same system but where the vacua $B_\pm$ now appear on the $\phi_1$-axis. Recall that these models admit a family of topological kinks only for the regimes $\beta>0$ and $\gamma > 0$, respectively, and hence, these solutions are related by the aforementioned transformation. 
\end{itemize}
In conclusion, it can be stated that the only models with a fourth degree polynomial potential with even symmetry in the scalar fields derived from polynomial superpotentials that yield families of kinks correspond to BNRT-type models and their possible variations through transformations. 

\section{Models admitting superpotentials with one singular point}

As previously noted, alternative forms of the superpotential $W(\phi_1,\phi_2)$ can be chosen while still generating a fourth-degree polynomial potential $U(\phi_1,\phi_2)$, which remains the primary focus of this work. In particular, one may consider superpotentials involving irrational functions. Furthermore, if these functions incorporate a critical point $(a_1,a_3)$, where the function is not differentiable, the possibility of finding families of semi-BPS kinks arises. For the purposes of this section, we shall adopt the following generic form for the superpotential:
\begin{equation}
    W(\phi_1,\phi_2)=\sqrt{(\phi_1-a_1)^2+a_2\,(\phi_2-a_3)^2}\,\Big(b_1\,\phi_1^2+b_2\,\phi_2^2+b_3\,\phi_1\,\phi_2+b_4\,\phi_1+b_5\,\phi_2+b_6\Big);\qquad a_i,b_j\in\mathbb{R}.
    \label{eq:SUPRAD}
\end{equation}
Substituting this expression again into \eqref{eq:potential02} and imposing the form \eqref{eq:POTG} and the $\mathbb{Z}_2\times\mathbb{Z}_2$ symmetry, we obtain the minimal set of independent parameters that generate a fourth degree polynomial potential. Two one-parameter families of models can be identified in this case, which are listed below:

\begin{itemize}
    \item \textbf{Case 3:} \textit{MSTB type models}: In this case the superpotential is given by the expression
\begin{equation}
    W_3(\phi_1,\phi_2)=\frac{1}{3}\sqrt{(\phi_1\pm\sigma)^2+\phi_2^2}\,\Big(\phi_1^2+\phi_2^2\mp\sigma\,\phi_1+\sigma^2-3 \Big),\label{eq:MSTBW}
\end{equation}
 which defines the family of the so called MSTB potentials     
\begin{equation}
    U_3(\phi_1,\phi_2)=\frac{1}{2}(1-\phi_1^2)^2+ \frac{1}{2}(1-\phi_2^2)^2+\phi_1^2\,\phi_2^2+\frac{\sigma^2}{2}\,\phi_2^2-\frac{1}{2},\label{eq:MSTB}
\end{equation}
where $\sigma\in\mathbb{R}$. It should first be noted that \eqref{eq:MSTBW} defines two distinct families of superpotentials, each characterized by the choice of sign in the ($\pm$) expression found in \eqref{eq:MSTBW}. The singular point of the superpotential is different in each case, namely $F_\pm=(\pm \sigma, 0)$. In other words, for a given value of the coupling constant $\sigma$, each potential \eqref{eq:MSTB} possesses two different superpotentials. Consequently, this leads to two (non-trivial) systems of first-order equations that govern the kink-type solutions and the possibility of the existence of two kink families. For the sake of completeness we recall here the structure of its kink variety:

\begin{enumerate}
    \item If $\sigma^2 \geq 1$, the kink manifold reduces to the existence of the single kink \eqref{TK1}. Therefore, no continuous families of solutions are present in this case and for this reason the kink \eqref{TK1} is stable in this regime.

    \item Conversely, for $0<\sigma^2<1$, it can be shown that there exist three topological kinks/antikinks described by the expressions \eqref{TK1} and 
    \begin{equation}
    \mathcal{K}^{(\alpha,\beta)}_{A\overline{A}}(x)=\left((-1)^{\alpha}\,\tanh\sigma\bar{x},(-1)^{\beta}\,\sqrt{1-\sigma^2}\,\text{sech}\,\sigma\bar{x}\right),
    \label{eq:MSTBTK2}
\end{equation}
    where $\alpha,\beta=0,1$ and $\bar{x}\equiv x-x_0$, being $x_0\in\mathbb{R}$ an integrating constant that represents the center of the kink. In this regime the one non-null component kinks \eqref{TK1} become unstable and decay to the stable solution \eqref{eq:MSTBTK2}, which has less energy,
    \begin{equation}
        E[\mathcal{K}_{A\overline{A}}(x)]=2\,\sigma\left(1-\frac{\sigma^2}{3}\right)\,.
    \end{equation}
    The orbit of these solutions conforms the ellipse $\phi_1^2+\phi_2^2/(1-\sigma^2)=1$, whose foci correspond to the singular points of the superpotentials (\ref{eq:MSTBW}), $F_{\pm}\equiv(\pm\sigma,0)$. The orbits of these topological solutions are depicted in Figure \ref{fig:MSTBTK}, and their profiles, in Figure \ref{fig:MSTBprof}. Observing the energy densities in Figure \ref{fig:MSTBprof} we can see that both the $\mathcal{K}_1(x)$ kink and the $\mathcal{K}_{A\overline{A}}(x)$ kink are the basic particles of the model.
    
  \begin{figure}[h]
  \centering
    \includegraphics[width=6.5cm]{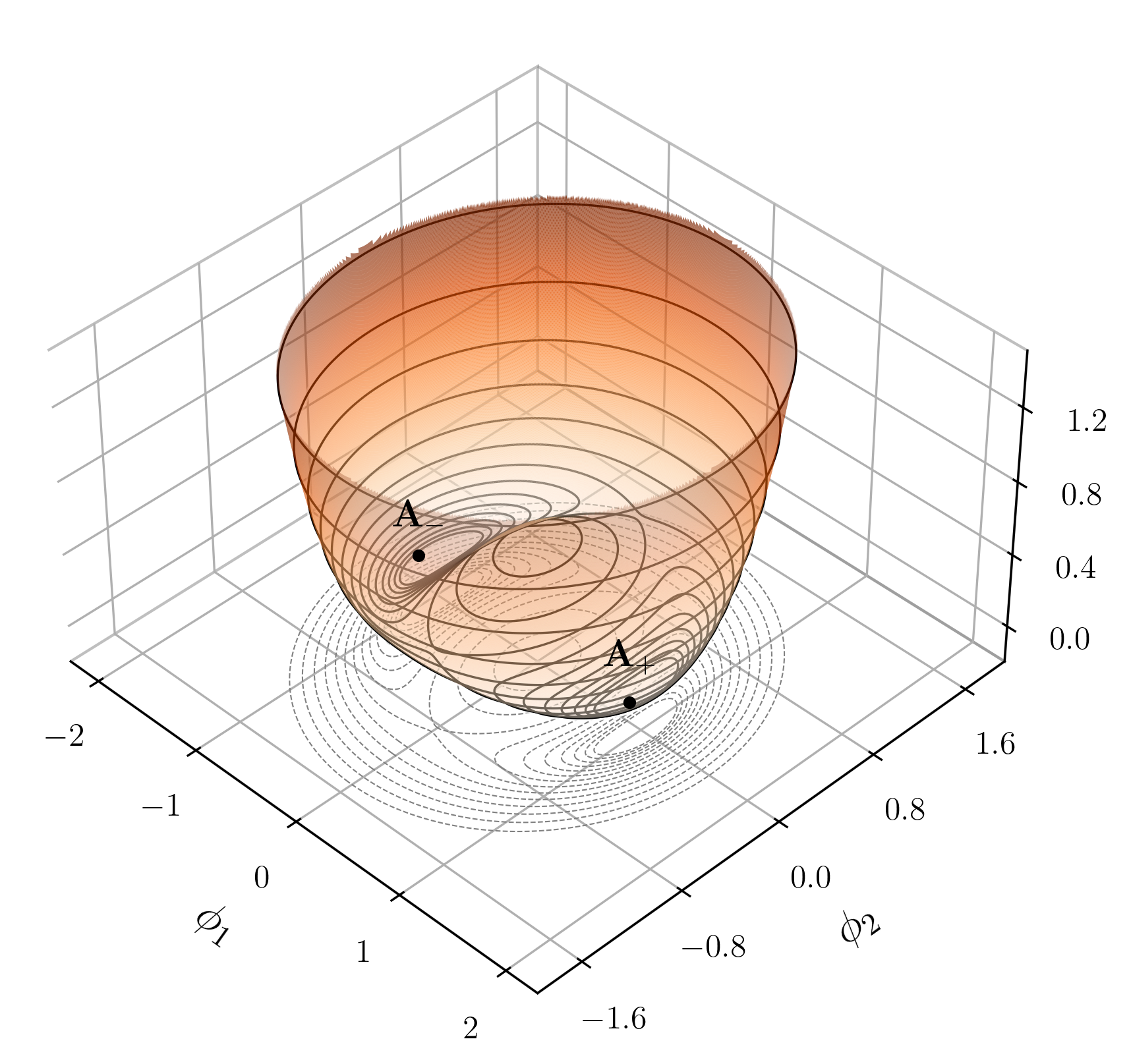}
    \includegraphics[width=6.5cm]{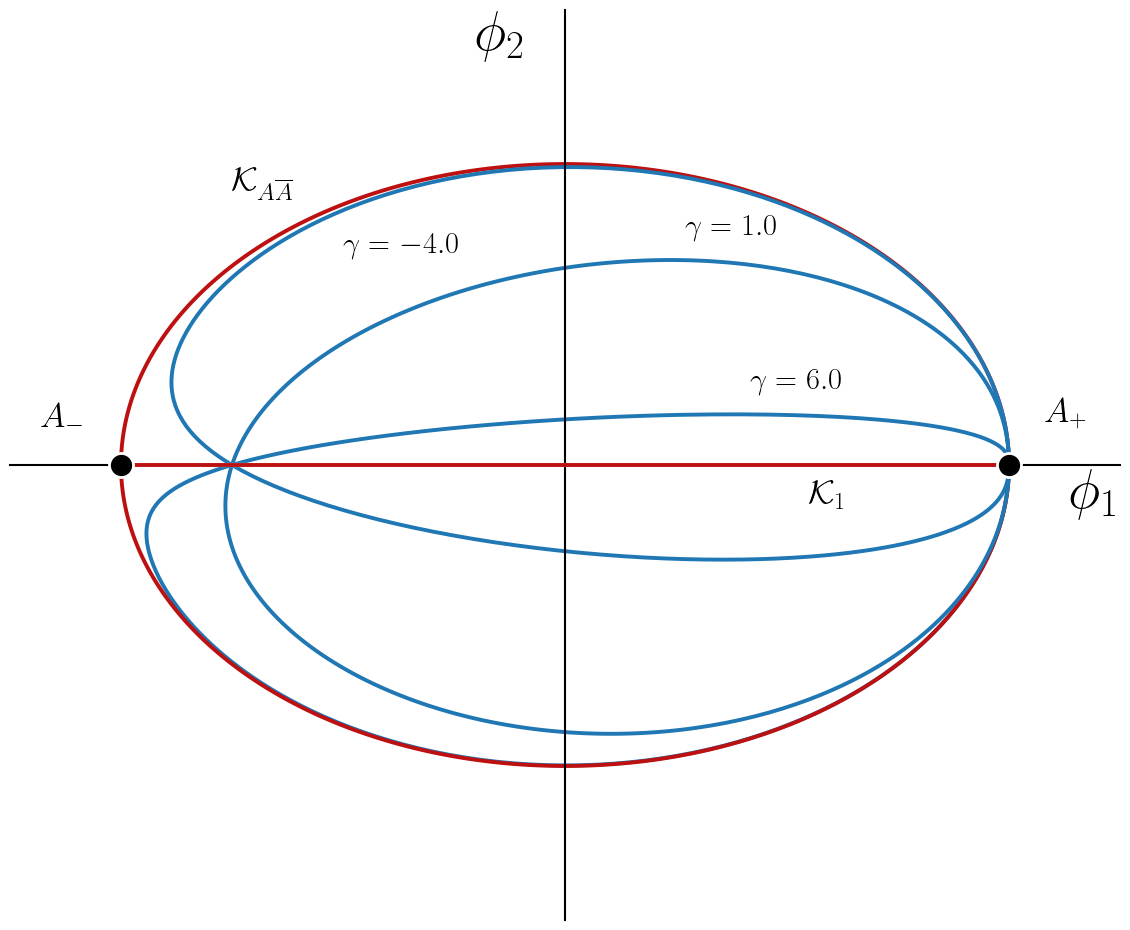} 
  \caption{Representation of the potential (left) and the orbits (right) of the MSTB model for $\sigma = 3/4$. Here, the red curves are the topological kinks of the model, $\mathcal{K}_{A\overline{A}}(x)$ and $\mathcal{K}_1(x)$. The blue curves correspond to the non-topological solutions $\mathcal{K}_{AA}(x;\gamma)$, \eqref{eq:MSTBNTK}, for several values of the family parameter $\gamma$.}
  \label{fig:MSTBTK}
   \end{figure}
    
Additionally, two families of non-topological kinks emerges, following the expression 
\begin{align}
    \mathcal{K}_{AA}^{(\alpha)}(x;\gamma) = \left((-1)^{\alpha}\frac{\sigma_-\cosh\sigma_+x_+-\sigma_+\cosh\sigma_-x_-}{\sigma_-\cosh\sigma_+x_++\sigma_+\cosh\sigma_-x_-},\frac{\sigma_+\sigma_-\sinh\bar{x}}{\sigma_-\cosh\sigma_+x_+-\sigma_+\cosh\sigma_-x_-}\right).
    \label{eq:MSTBNTK}
\end{align}
Here $\gamma$ is the parameter that identifies each member of the family and we have denoted $x_{\pm}\equiv \bar{x}-\gamma\sigma(1\mp\sigma)$ and $\sigma_{\pm}=1\pm\sigma$. These latter solutions asymptotically depart from the vacuum point $A_\pm$ and return (asymptotically) to it without reaching a different vacuum, all of them passing through the singular point $F_\mp$. The two kink families are related by the reflection transformation $\phi_1\to -\phi_1$. Notice that, by checking the energy density plot in Figure \ref{fig:MSTBprof} right, these non-topological kinks are composite solutions of two lumps, which can be suplemented by the energy sum-rule
\begin{equation}
    E[\mathcal{K}_{AA}(x;\gamma)] =E[\mathcal{K}_1(x)] + E[\mathcal{K}_{A\overline{A}}(x)].
\end{equation}
\end{enumerate}

The MSTB model is, indeed, a very special case. The dynamical system characterizing the static solutions of the field model is completely integrable, featuring two first integrals that can be explicitly calculated. Furthermore, the system is Hamilton-Jacobi separable; specifically, the associated second-order ordinary differential equations can be decoupled and solved by employing an elliptical coordinate system, where the isocoordinate curves have $F_\pm$ as their focal points. In other words, all such models are completely integrable in the aforementioned sense, regardless of the value of the coupling constant $\sigma$. This, in turn, explains the existence of two distinct superpotentials, which give rise to two sets of first-order ordinary differential equations. These equations can be derived by solving for the spatial derivatives from the first integrals of the associated mechanical problem and imposing the corresponding asymptotic conditions for kink-type solutions.

\begin{figure}[]
  \centering
    \includegraphics[width=5cm]{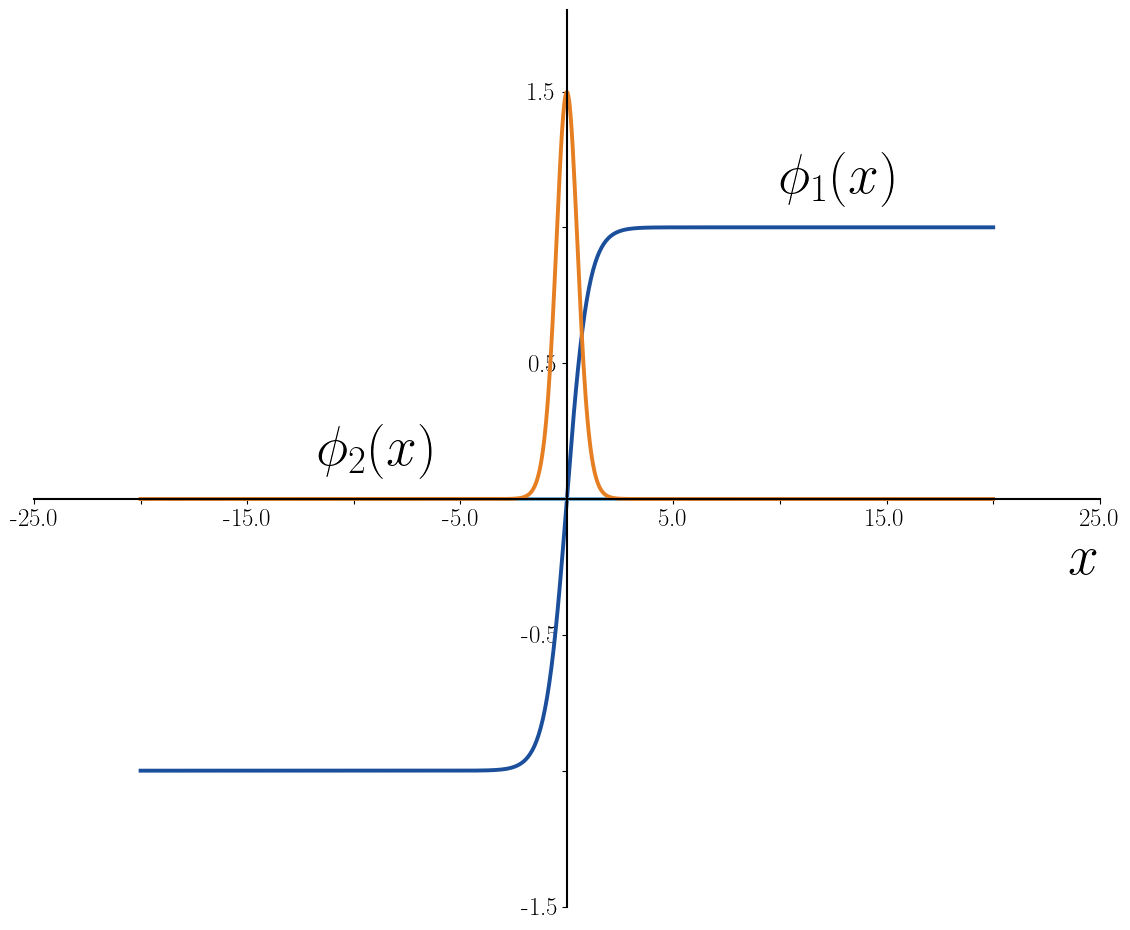} 
    \includegraphics[width=5cm]{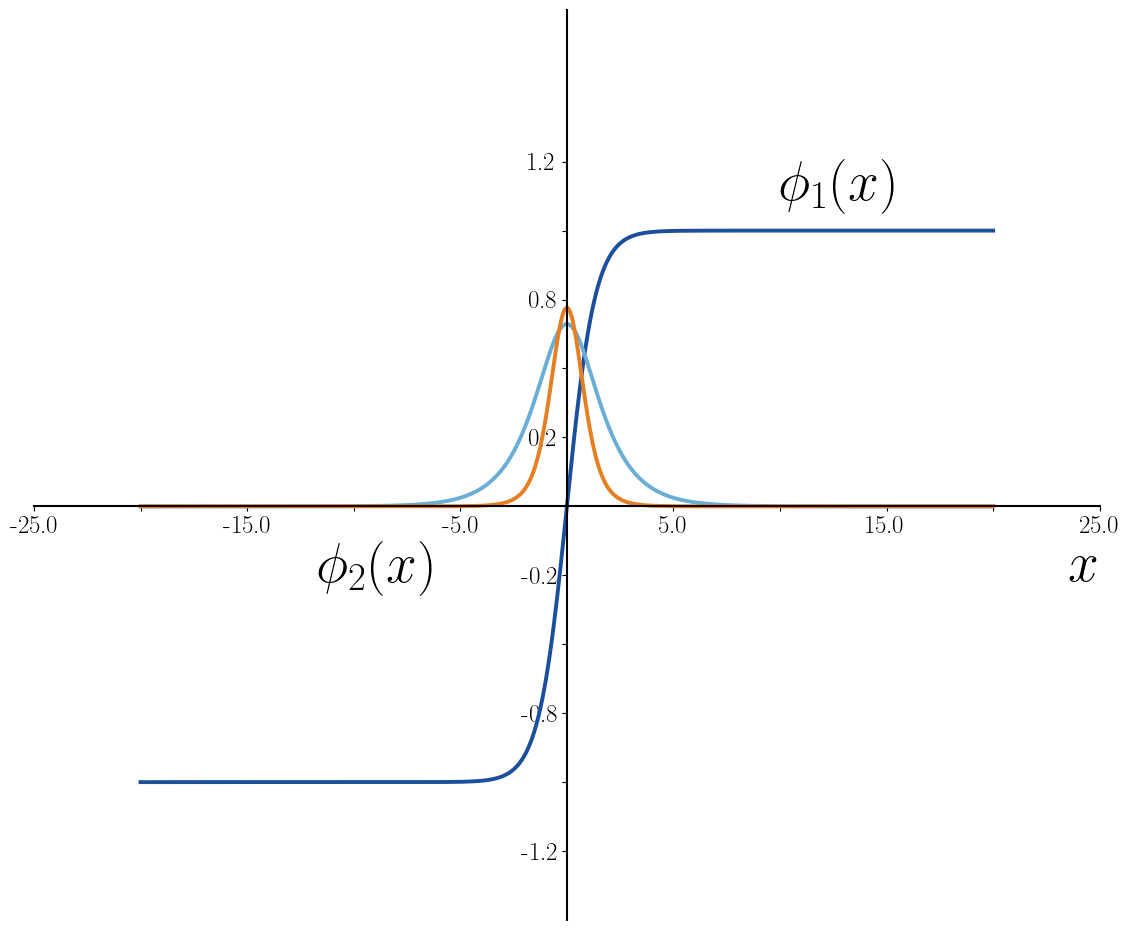}
    \includegraphics[width=5cm]{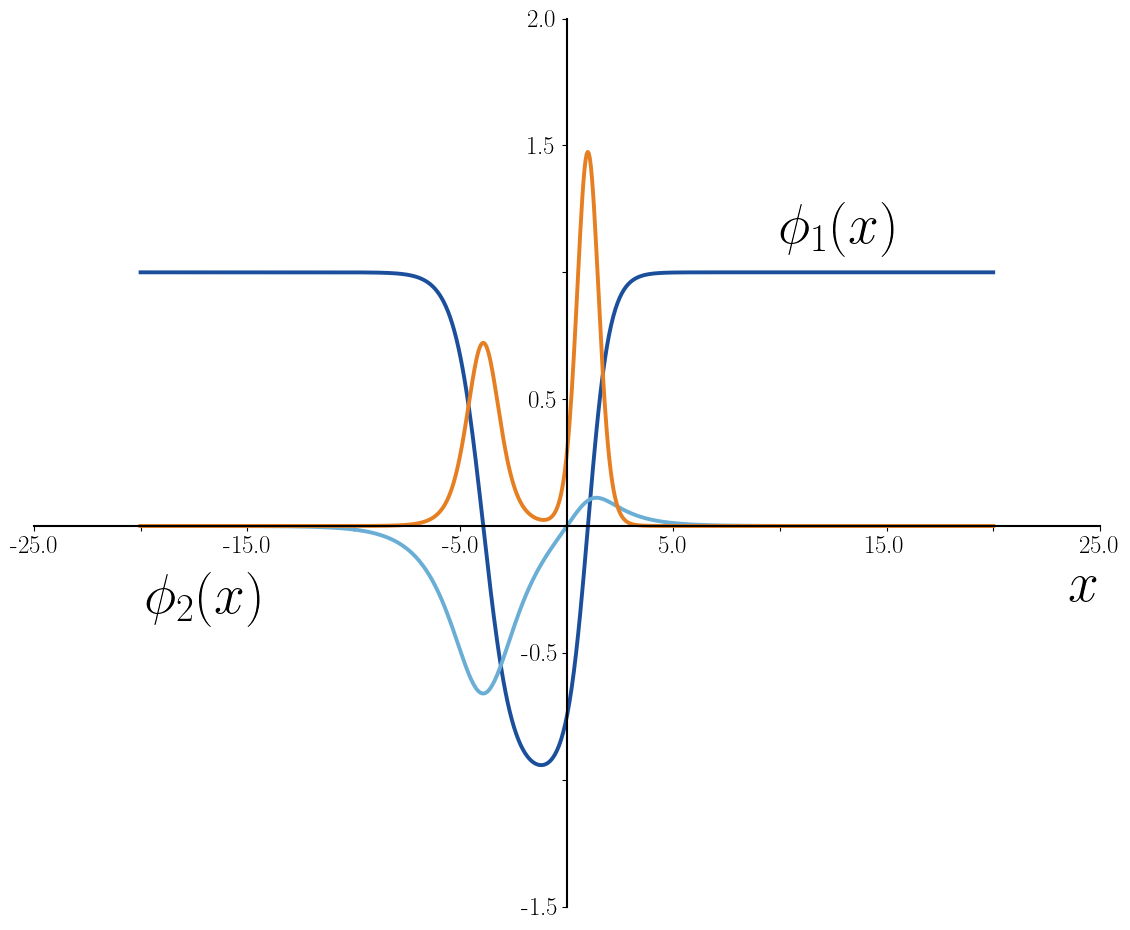}
  \caption{Graphical representation of the profiles $\Phi(x)$ of the MSTB model (dark blue for $\phi_1(x)$ and light blue for $\phi_2(x)$) and their associated energy densities (orange curve). From left to right: $\mathcal{K}_1(x)$, $\mathcal{K}_{A\overline{A}}(x)$ and $\mathcal{K}^{(1)}_{AA}(x;6)$.}
  \label{fig:MSTBprof}
\end{figure}

\item \textbf{Case 4:} \textit{A new family of models with fourth-degree polynomial potential:} Alongside the previous case, we identify a completely novel model that, to the best of our knowledge, has not been addressed in the literature. This constitutes the primary result of the present work: the identification of a new model featuring a fourth-order polynomial potential which possesses kink families. For this case, the superpotential is given by the expression: 
\begin{equation}
     W_4(\phi_1,\phi_2)=\frac{1}{3}\sqrt{\phi_1^2+\phi_2^2}\,(\phi_1^2+\mu\,\phi_2^2-3);\label{eq:WRAD}
\end{equation}
 which gives place to the one-parameter family of potentials
\begin{equation}
    U_4(\phi_1,\phi_2)=\frac{1}{2}(1-\phi_1^2)^2+\frac{1}{2}(1-\mu\,\phi_2^2)^2+\frac{1}{9}(2+\mu)(1+2\mu)\,\phi_1^2\,\phi_2^2-\frac{1}{2},\label{eq:POTR1}
\end{equation}
where $\mu\in\mathbb{R}$ is the coupling constant. For this reason, we shall perform a more detailed analysis of this model, identifying the variety of kink-type solutions and explicitly discussing their stability whenever possible. The static field equations associated with \eqref{eq:POTR1} are
\begin{align}
    & \frac{d^2\phi_1}{dx^2}=\frac{2}{9} (\mu +2) (2 \mu +1) \,\phi _1 \,\phi _2^2+2 \,\phi _1
   \left(\phi _1^2-1\right)\label{eq:FEP2}\\
    & \frac{d^2\phi_2}{dx^2}=\frac{2}{9} (\mu +2) (2 \mu +1) \,\phi _1^2\, \phi _2-2 \,\mu \, \phi
   _2 \left(1-\mu \, \phi _2^2\right)\,\,.\nonumber
\end{align}
It can be verified that for the case $\mu=1$, the potential (\ref{eq:POTR1}) is rotationally invariant, giving rise to a continuous set of vacua localized on the unit circle $S^1=\{(\phi_1,\phi_2)\in \mathbb{R}^2: \phi_1^2 + \phi_2^2=1\}$. By employing polar coordinates, this case can be reduced to a model where only the radial coordinate in the internal field space plays a relevant role; consequently, the results reduce to the standard kink (\ref{TK1}), which can be rotated within the internal plane. Excluding this particular case, the model can be considered to exhibit two distinct regimes, Figure \ref{fig:Vnew}:
  \begin{figure}[h]
  \centering
    \includegraphics[width=6.5cm]{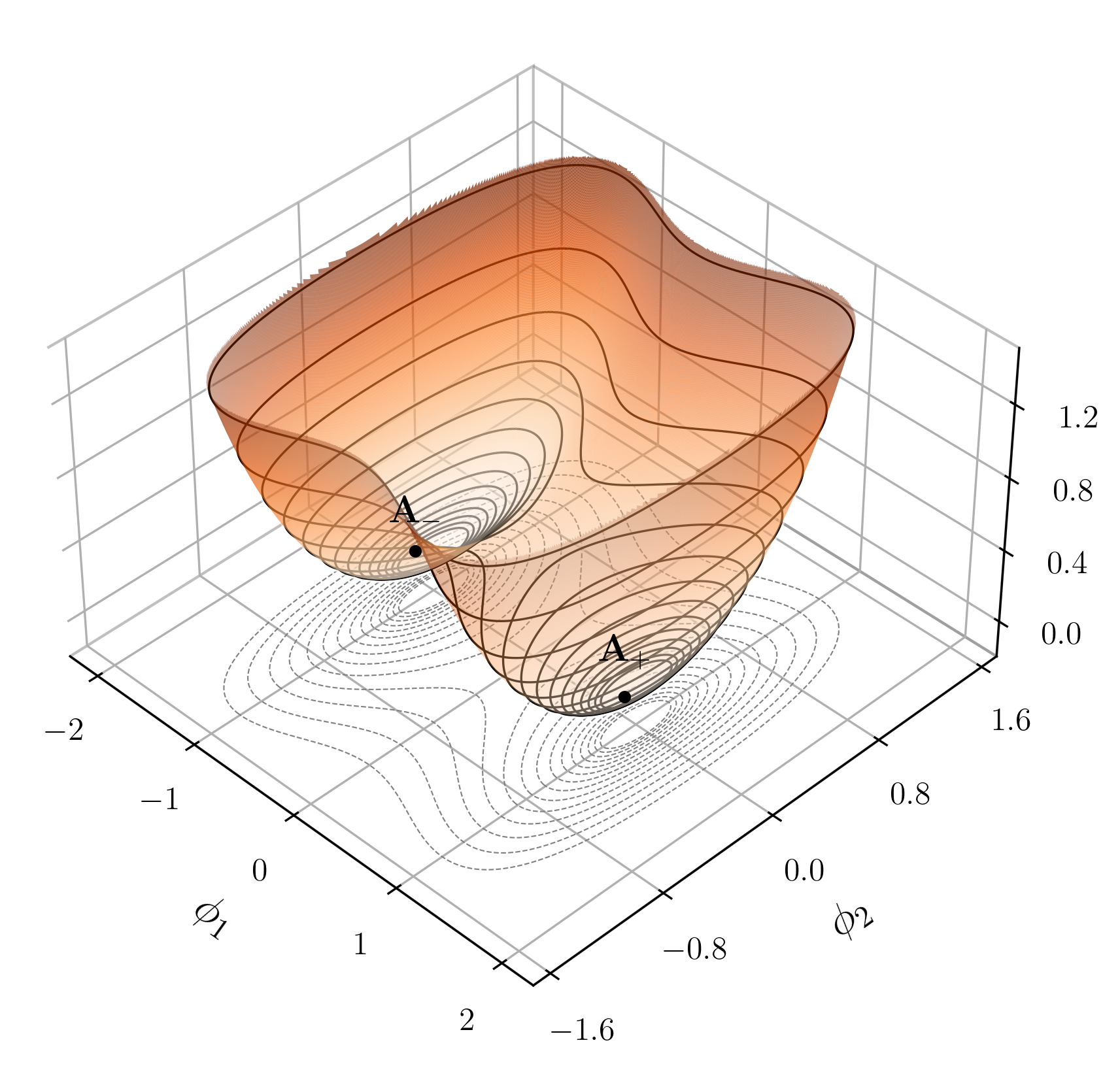}
    \includegraphics[width=6.5cm]{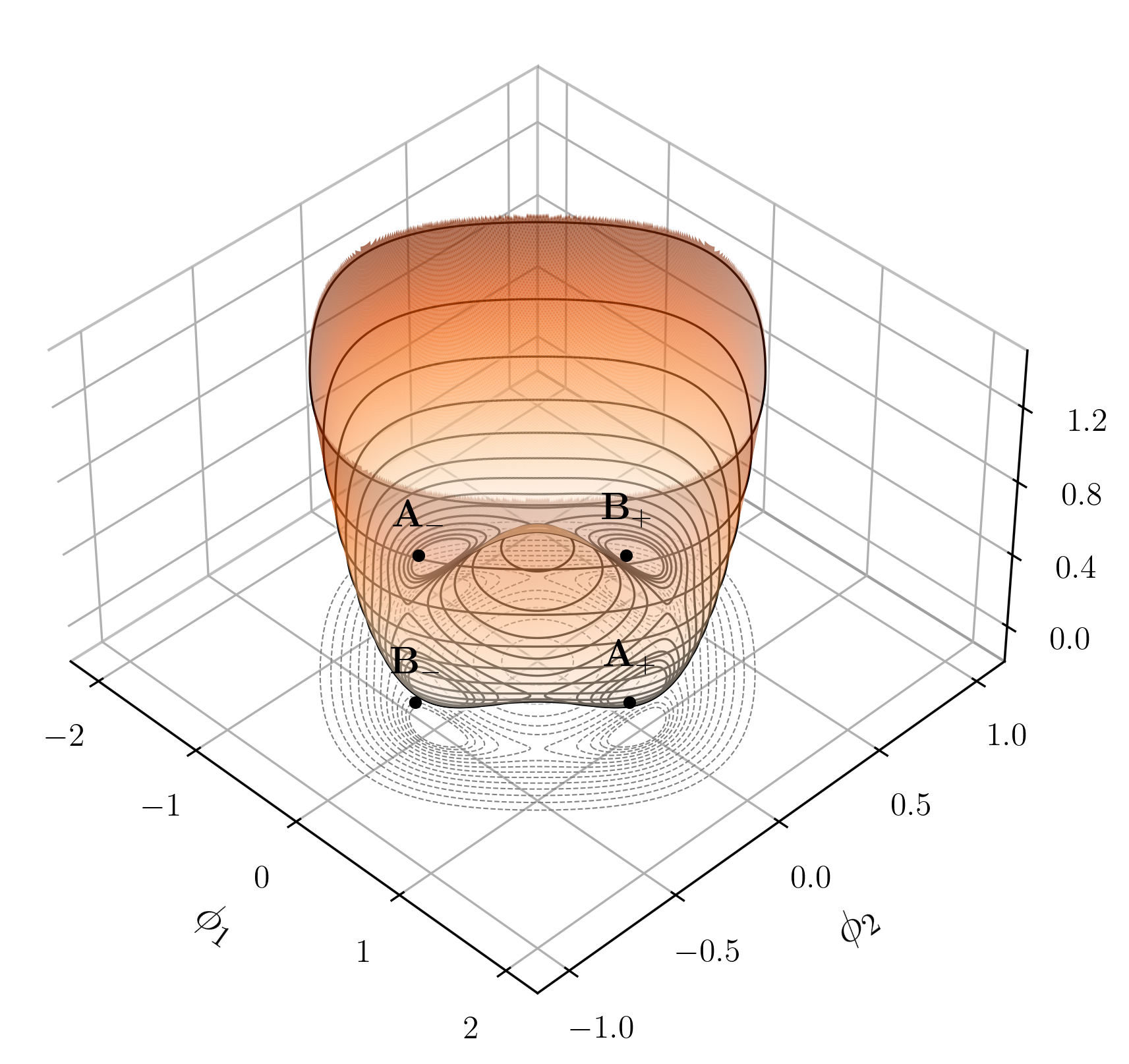} 
  \caption{Representation of the potential $U_4$ for $\mu=-1/2$ (left) and $\mu=4$ (right).}
  \label{fig:Vnew}
   \end{figure}
\vspace{0.2cm}

$\bullet$ If $\mu<0$ the potential only involves two vacua, which are located at $A_\pm$ as previously prescribed. In this case, the kink manifold is restricted to the solution (\ref{TK1}). As usual, the study of the linear stability of this solution reduces to the spectral problem of the second-order small fluctuations operator valued on the previous solution, which in this instance is given by
\begin{equation}
    \mathcal{H}[\mathcal{K}_1^{(\alpha)}(x)]=\begin{pmatrix}
        & -\frac{d^2}{dx^2}+4-6\,\text{sech}^2\bar{x}& 0 \\
        & 0 & -\frac{d^2}{dx^2}+\frac{4}{9}(\mu - 1)^2-\frac{2}{9}(2+\mu)(1+2\mu)\,\text{sech}^2\bar{x}\,\,
    \end{pmatrix}. \label{hess4}
\end{equation}
Longitudinal perturbations are those of the $\phi^4$ kink and will not be commented again.      Regarding the contributions from orthogonal perturbations, it is clear that the potential well associated with the operator (\ref{hess4}) takes values in the range between $2|\mu|$ and $\frac{4}{9}(\mu-1)^2$. In other words, the potential is strictly positive throughout this regime. Consequently, the eigenvalues of the operator (\ref{hess4}) are necessarily positive, which implies the stability of the kink solution $\mathcal{K}_1(x)$ for any negative value of $\mu$.

\vspace{0.2cm}

$\bullet$ The $\mu>0$ case is more interesting since now there are four vacua
\begin{equation}
    \mathcal{M}=\left\{ A_{\pm}=(\pm1,0)\hspace{0.2cm}, \hspace{0.4cm} B_{\pm}=(0,\pm1/\sqrt{\mu}) \right\} .
\end{equation}
and a family of semi-BPS kinks arises in this regime. By construction, there exist kink-type solutions that must satisfy the gradient equations derived from the superpotential (\ref{eq:WRAD}),
\begin{align}
&\frac{d\phi_1}{dx}=\pm\,\frac{\phi _1 \left((\mu +2) \,\phi _2^2+3\, \phi _1^2-3\right)}{3 \sqrt{\phi _1^2+\phi _2^2}}\label{eq:BOGPP2}\\
&\frac{d\phi_2}{dx}=\pm\,\frac{\phi _2 \left((2 \mu +1)\, \phi _1^2+3 \,\mu  \,\phi _2^2-3\right)}{3 \sqrt{\phi _1^2+\phi _2^2}}.\nonumber
\end{align}
The singular point of \eqref{eq:WRAD} becomes a singularity of the equations. The orbit equation derived from (\ref{eq:BOGPP2})
    \begin{equation}
        \frac{d\phi_2}{d\phi_1}=\frac{\phi _2 \left((2 \,\mu +1)\, \phi _1^2+3\, \mu\,  \phi
   _2^2-3\right)}{\phi _1\, \left((\mu +2)\, \phi _2^2+3\, \phi
   _1^2-3\right)}
   \label{eq:BOGOorb}
    \end{equation}
    can be solved by using the integrating factor
    \begin{equation}
        \mu(\phi_1,\phi_2)=\frac{ \phi _1^3  \left((\mu +2)\, \phi _2^2+3 \phi
   _1^2-3\right)}{(1-\mu )\, \phi _2 \left(\phi _1^2+\phi _2^2\right){}^{3/2}}\left(\frac{2(1-\mu )\, \phi _2^2}{\phi _1^2}\right){}^{\frac{3}{2-2 \mu }},
    \end{equation}
    in the previous equation. The kink orbits in this case can be written as
    \begin{equation}
        F(\phi_1,\phi_2;\kappa)\equiv\phi_1^2-\sqrt{1+z}\,\,z^{\frac{-3}{2(1-\mu)}}\Bigg(\kappa +(-1)^{\frac{-3}{2(1-\mu)}}{\frac{3}{2(1-\mu)}}B_{-z}\Big(\frac{3}{2(1-\mu)},\frac{-1}{2}\Big)\Bigg)=0;
        \label{eq:ORBSOL}
    \end{equation}
    where $\kappa\in\mathbb{R}$ is an integrating constant that labels every member of the family; $z\equiv\frac{\phi_2^2}{\phi_1^2}$, and $B_u(a,b)$ is the incomplete Euler Beta function. Note that $A_+$ and $A_-$ are minima of $W_4$ for $0<\mu <1$ and saddle points otherwise. Conversely, $B_+$ and $B_-$ are minima for $\mu\in(0,1)$ and saddle points if $\mu >1$. As a result, we expect to find topological solutions connecting $B_+$ and $B_-$ for $\mu \in(0,1)$ and linking $A_+ $ with $A_-$, for $\mu>1$. Nevertheless, flow lines of $\text{grad}\, W$ between one point in the $AA$ sector and another point in the $BB$ sector exist for every $\mu>0$. These remarks suggest the convenience of distinguishing between two different regimes in this model, characterized by either $0<\mu<1$ or $\mu>1$. 

    \begin{itemize}
        \item If $0<\mu<1$ the orbits (\ref{eq:ORBSOL}) with $\kappa < 0$ determine families of curves that asymptotically starts at the vacua $B_+$ or $B_-$, remain within a single quadrant of the internal plane, and converge toward the origin. All such orbits pass through the origin before crossing into the opposing quadrant, eventually reaching the $B_-$ or $B_+$ vacuum, respectively, see Figure \ref{fig:NuevOrb} (left). We will refer to these solutions as $\mathcal{K}_{B\overline{B}}(x,\kappa)$. Note that as a member of the family $\mathcal{K}_{B\overline{B}}(x,\kappa)$ we can find a one non-null component kink determined by
        \begin{equation}
        \mathcal{K}_2^{(\alpha)}(x)=\Big(0,\frac{(-1)^{\alpha}}{\sqrt{\mu}}\,\tanh\sqrt{\mu}\,\bar{x}\Big). \label{TKY}
        \end{equation}
        for the value $\kappa\to\infty$. 
        
        For $\kappa=0$, (\ref{eq:ORBSOL}) yields isolated curves that connect one of the vacuum points $A_+$ or $A_-$ with one of the vacua $B_+$ or $B_-$. In total, these correspond to eight isolated kinks and antikinks, which will be denoted by $\mathcal{K}_{AB}(x)$. By using (\ref{enegia2}) it can be calculated the energy of these solutions,
        \[
        E[\mathcal{K}_{B\overline{B}}(x,\kappa)]=\frac{4}{3\sqrt{\mu}} \hspace{0.5cm},\hspace{0.5cm}  E[\mathcal{K}_{AB}(x)]= \frac{2}{3}\left(\frac{1}{\sqrt{\mu}}-1\right)
        \]
        Recall that in addition to the family $\mathcal{K}_{B\overline{B}}(x,\kappa)$ (obtained from the first order equations (\ref{eq:BOGPP2})) we can find for any value of the coupling constant $\mu$ the kink (\ref{TK1}), in this occasion as a solution of the second order equations (\ref{eq:FEP2}). The total energy in this case is
        \[
        E[\mathcal{K}_1(x)]=  \frac{4}{3}
        \]
        Therefore, the next energy sum rule
        \[
        E[\mathcal{K}_{B\overline{B}}(x,\kappa)] = 2 \, E[\mathcal{K}_{AB}(x)] + E[\mathcal{K}_1(x)]
        \]
        holds. This means that kinks in the sector $BB$ in this regime can be interpreted as composite extended particles comprising two more basic energy lumps associated to the kinks $\mathcal{K}_{AB}(x)$ and one lump associated to the single kink $\mathcal{K}_1(x)$.

        On the other hand, it should be noted that the configuration in the sector $AA$ comprising two $\mathcal{K}_{AB}(x)$-kinks is energetically more favorable than the $\mathcal{K}_1(x)$-kink if $\mu>\frac{1}{4}$, where as it becomes more energetic when $\mu<\frac{1}{4}$. Naturally, this feature has significant implications for the stability of these configurations, as will be discussed later.
        
        \item On the other hand, if $\mu>1$ we find a pattern similar to the one described in the previous case; however, the one-parameter family of kink solutions now asymptotically connects the vacuum points $A_-$ and $A_+$, with each member of the family passing through the origin, see Figure \ref{fig:NuevOrb} (right). Consequently, this family shall be referred to as $\mathcal{K}_{A\overline{A}}(x,\kappa)$. This behavior occurs for values of $\kappa$ in the range $\kappa<\kappa_c$ with
        \[
        \kappa_c = -\frac{2}{\sqrt{\pi}}\Gamma\Big(1+\frac{3}{2(1-\mu)}\Big)\Gamma\Big(\frac{3}{2}-\frac{3}{2(1-\mu)}\Big)\,,
        \]
  
        being $\Gamma(u)$ the Euler Gamma function. If the constant $\kappa$ reaches the critical value $\kappa_c$, isolated kink solutions emerge within the $AB$ topological sector; therefore, we once again employ the notation $\mathcal{K}_{AB}(x)$ to denote them. In this case, the kink energies are given by
        \[
        E[\mathcal{K}_{A\overline{A}}(x,\kappa)]=\frac{4}{3} \hspace{0.5cm},\hspace{0.5cm}  E[\mathcal{K}_{AB}(x)]= \frac{2}{3}\left(1-\frac{1}{\sqrt{\mu}}\right)\,.
        \]
        In this case, the kink $\mathcal{K}_1(x)$ given by the expression (\ref{TK1}) is a member of the family $\mathcal{K}_{A\overline{A}}(x,\kappa)$ with $\kappa\to\infty$ and the single solution $\mathcal{K}_2(x)$ in (\ref{TKY}) verifies the second order equation (\ref{eq:FEP2}) with a total energy
        \[
        E[\mathcal{K}_{2}(x)] = \frac{4}{3\sqrt{\mu}}
        \]
        leading to the energy sum rule
        \[
        E[\mathcal{K}_{A\overline{A}}(x,\kappa)] = 2 \, E[\mathcal{K}_{AB}(x)] + E[\mathcal{K}_2(x)]
        \]
        Note that in this regime the kink $\mathcal{K}_2(x)$ is less energetic than the kink $\mathcal{K}_{A\overline{A}}(x,\kappa)$, such that the previous relation is coherent. 
     \end{itemize}

  \begin{figure}[h!]
  \centering
    \includegraphics[width=6.5cm]{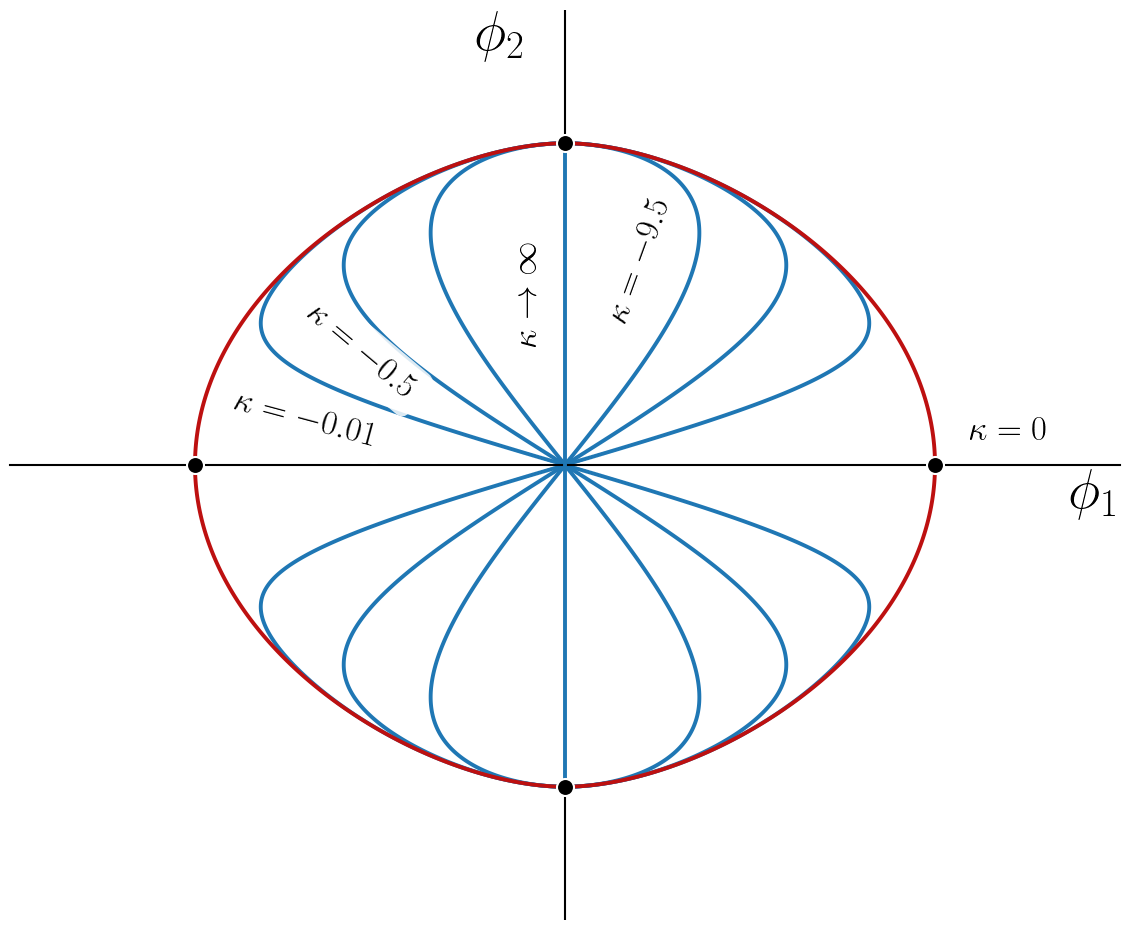} \hspace{1cm}
    \includegraphics[width=6.5cm]{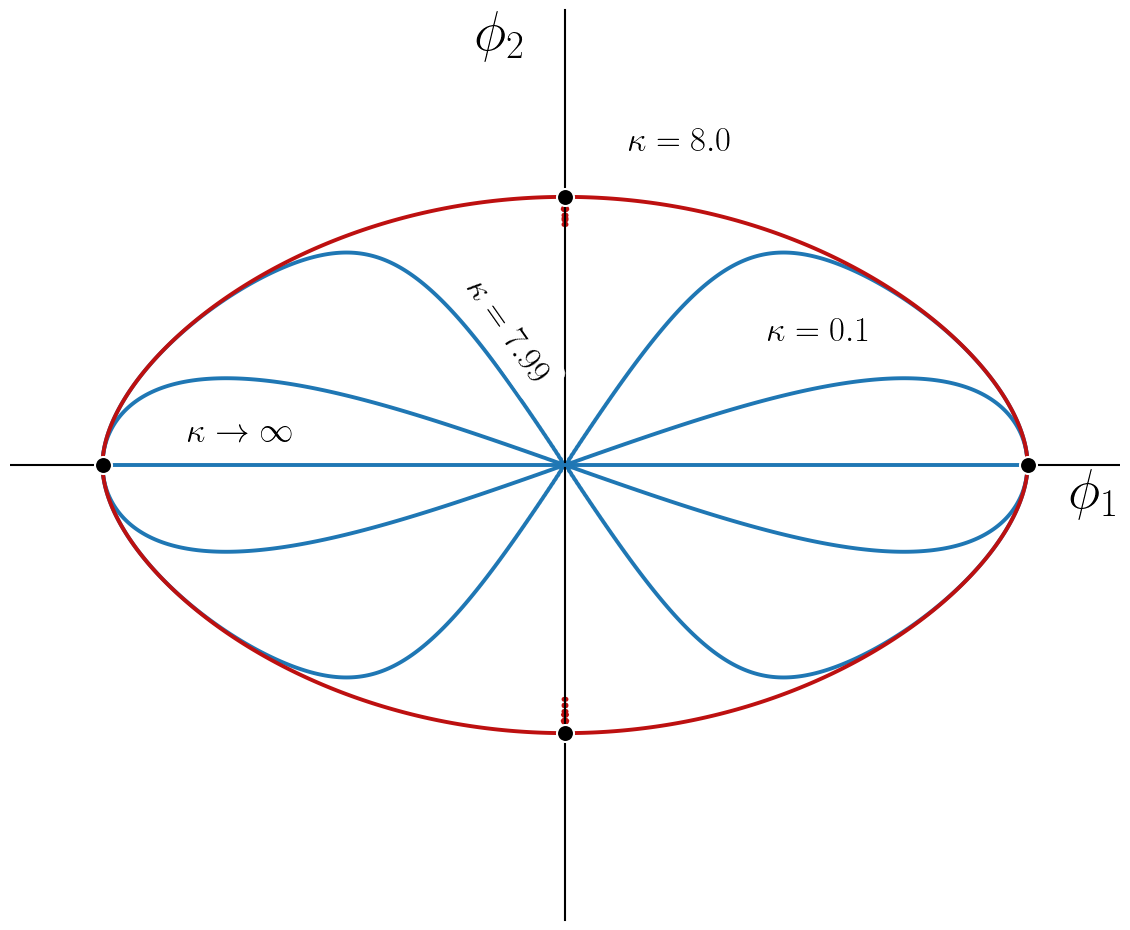}
  \caption{Graphical representation of the orbits \eqref{eq:ORBSOL} in the internal space for $\mu = 1/2$ (left) and $\mu = 2$ and several values of $\kappa$. The black dots represent the vacua. The red curves correspond to the critical orbits, $\kappa_c(1/2)=0$ (left) and $\kappa_c(2) = 8$ (right).}
  \label{fig:NuevOrb}
\end{figure}

    Remarkably, the two previous regimes can be related by an internal symmetry. The model \eqref{eq:POTR1} is invariant under the internal transformation and reparameterization
    \begin{equation}
       \mu\mapsto\,1/\mu,\qquad \phi_1\mapsto \,\sqrt{\mu}\,\phi_2,\qquad \phi_2\mapsto\,\sqrt{\mu}\,\phi_1;\qquad \mu >0.
       \label{eq:TRANS}
    \end{equation}
    Hence, we can identify the solutions for $\mu\in(1,\infty)$ from those in $\mu\in(0,1)$. 
    
 From the kink orbit flow, \eqref{eq:ORBSOL}, we can obtain $\phi_1=\phi_1(z)$ and therefore, using \eqref{eq:BOGPP2}, we could find the explicit dependence of $z$ on the spatial coordinate, $z(x)$, leading to the kink profile $\Phi(x)$. Nevertheless, the resulting equation for $z(x)$ cannot be integrated analytically, so in general $\Phi(x)$ cannot be computed in closed form. Some of these profiles have been obtained numerically, and their shapes are depicted in Figure \ref{fig:PROFGEN}, along with their energy densities. Here, it can be observed that, depending on the value of $\kappa$, the energy density presents one or three lumps, which indicates that the kinks in \eqref{eq:ORBSOL} are composite solutions where the separation distance is determined by the parameter $\kappa$. In fact, in Figure \ref{fig:PROFGEN}, for $\kappa = -9.5$ the particles are so close that the three lumps are superposed at the center of the kink.
    \begin{figure}[]
  \centering
    \includegraphics[width=5cm]{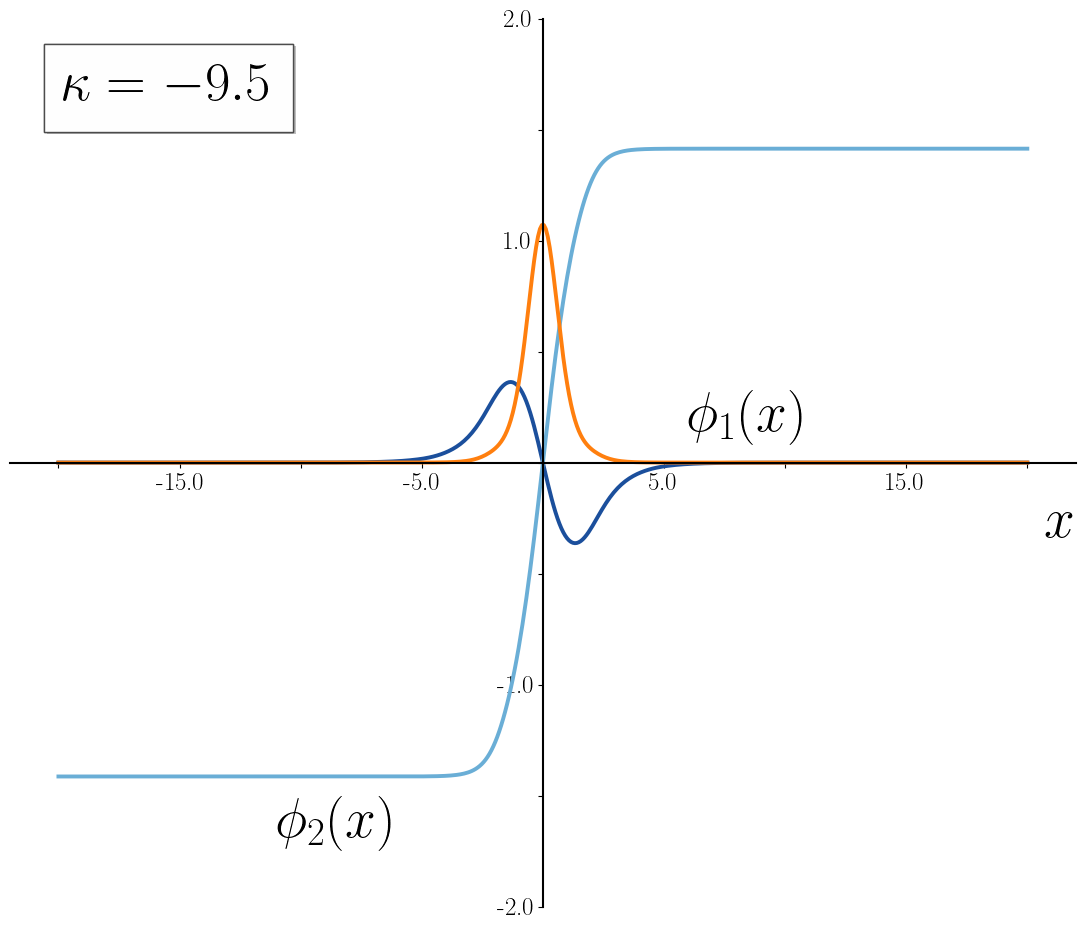} 
    \includegraphics[width=5cm]{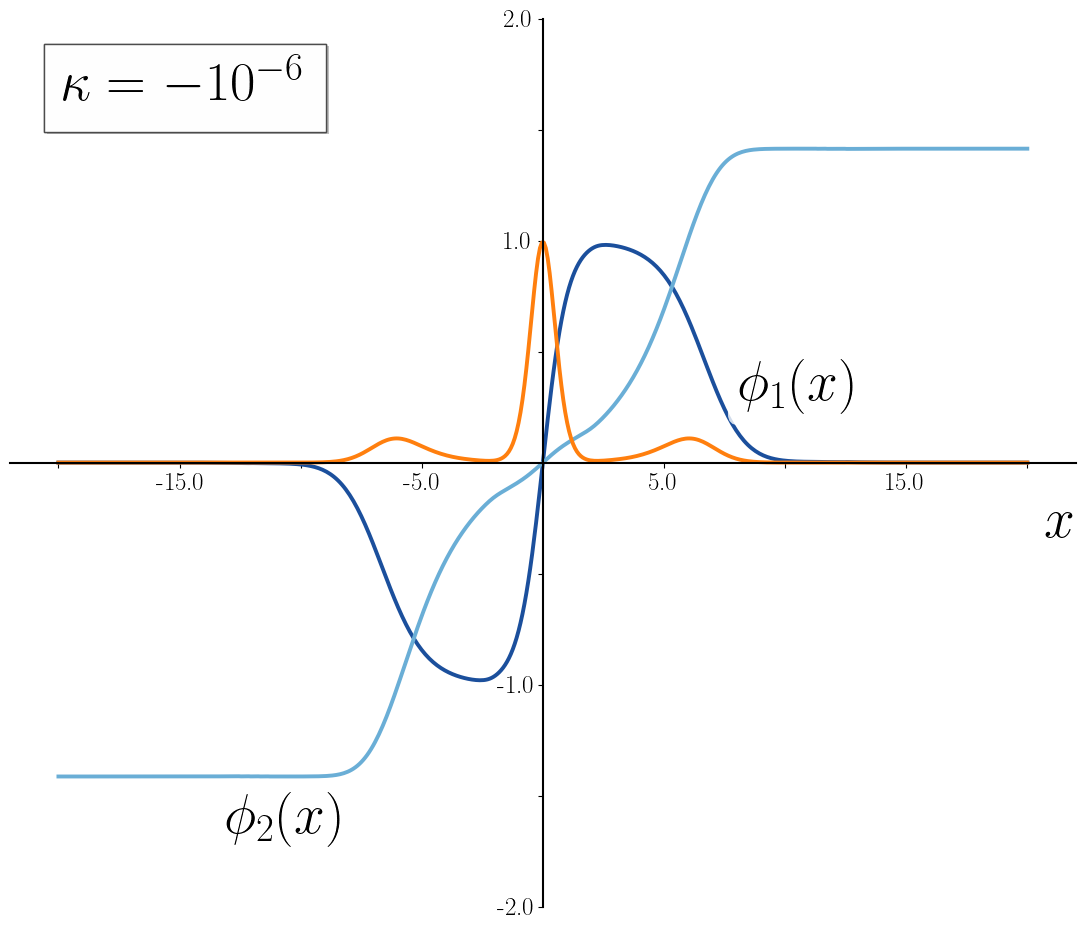}
  \caption{Graphical representation of the profiles $\Phi(x)$ (blue curves) obtained numerically for the orbits $\kappa=-9.5$ (left) and $\kappa=-10^{-6}$ (right) at $\mu = 1/2$ and their energy densities (orange curve).}
  \label{fig:PROFGEN}
\end{figure}

It is interesting to note that the one non-null component kinks $\mathcal{K}_1(x)$ and $\mathcal{K}_2(x)$ given by the expressions (\ref{TK1}) and (\ref{TKY}) continue to be solutions for any value of the coupling constant $\mu$, independently of the regimes where they arise as members of a kink family. We shall now take advantage of the simplicity of these solutions to analyze their stability. We will analyze the linear stability of the $\mathcal{K}_1(x)$ and $\mathcal{K}_2(x)$ kinks separately:

    \begin{itemize}
        \item \textit{Spectrum of small fluctuations around $\mathcal{K}_1(x)$}. As previously mentioned, the Hessian operator is given in this case by the operator (\ref{hess4}). The eigenvalues associated with the orthogonal fluctuations are given by:
        \[
        \omega_n^2 = -\frac{1}{3} (n-1) (3 \,n-4 \,\mu +1) \qquad \text{with}\qquad \,\,\,n=0,1,...,n < \text{E}[\textstyle\frac{2\,\mu+1}{3}]
        \]
        The lowest eigenvalue is
        \[
        \omega_0^2 = \frac{1}{3} (1-4\mu) 
        \]
        which it is negative if $\mu>\frac{1}{4}$, leading to a instability channel for this solution. Note that  in the regime $\mu > 1$, where the kink $\mathcal{K}_1(x)$ is a member of the family $\mathcal{K}_{A\overline{A}}$, the solution is therefore unstable. In the range $\frac{1}{4} < \mu <1$, where the $\mathcal{K}_1(x)$ is an isolated kink, this kink tends to decay to the configuration consisting of two $\mathcal{K}_{AB}$, more energetically favorable as we have previously commented.
        
        Besides in the regime $\mu > 1$, $n=1$ provides us with a null eigenvalue $\omega_{n=1}^2=0$, that is, it determines a second zero mode. Note that the existence of this mode demands the compliance of the condition $\frac{1}{3} (2\mu+1)>1$. In addition to this, if $\mu>\frac{13}{4}$ a shape mode (with positive eigenvalue) emerges. As $\mu$ grows new shape modes arise.

        On the other hand, if $0<\mu<\frac{1}{4}$ the discrete spectrum involves only the ground state, which has a positive eigenvalue and the kink $\mathcal{K}_1(x)$ is stable. Thus, a configuration formed by two $\mathcal{K}_{AB}$-kinks in the sector $AA$ could decay to this single kink because in this regime this last kink is less energetic than the previous configuration.

        \item \textit{Spectrum of small fluctuations around $\mathcal{K}_2(x)$.}  The kink fluctuation operator in this case is
        \begin{equation}
            \mathcal{H}[\mathcal{K}_2^{(\alpha)}(x)]=\begin{pmatrix}
                -\frac{d^2}{dx^2}+\frac{2 (\mu +2) (2 \mu +1) \left(1-\text{sech}^2\sqrt{\mu }\,\bar{x}
                \right)}{9 \mu }-2&  0\\ 
                0 & -\frac{d^2}{dx^2}+2 \mu  \left(2 -3\,\text{sech}^2\,\sqrt{\mu }\,\bar{x} \right) \\
        \end{pmatrix}. \label{Hessiano2}
        \end{equation}
        such that the orthogonal fluctuation spectrum is given by the eigenvalues
        \begin{equation}
            \omega_n^2=-\frac{1}{3} \, (n-1) (3 n \mu +\mu -4),\qquad n = 0,1,...,\text{E}[\textstyle\frac{\mu+2}{3\mu}].
        \end{equation}
        The lowest eigenvalue $\omega_{n=0}^2 = \frac{1}{3}(\mu-4)$ is negative if $\mu<4$ and positive for $\mu>4$. This means that the solution $\mathcal{K}_2(x)$ is unstable in the first case, which includes the regime $0<\mu<1$ where this solution is a member of the continuous kink family $\mathcal{K}_{B\overline{B}}(x,\kappa)$ and the range $1<\mu<4$ where the $\mathcal{K}_2(x)$ is an isolated kink. For $\mu>4$ this kink becomes stable. Note the symmetry with respect to the previous stability analysis for the $\mathcal{K}_1(x)$-kink. 
                
        The next eigenvalue for $n=1$ is always zero, that is, it determines the second zero mode provided that $\frac{\mu+2}{3\mu}>1$, that is, when $\mu\in(0,1)$ and the $\mathcal{K}_2(x)$-kink is part of a family. The next shape mode arises for $\omega_2^2=\frac{1}{3}(4-7\mu)$ when $0<\mu<\frac{2}{5}$, etc.

        On the other hand, if $\mu>4$, again, the discrete spectrum involves only a positive eivenvalue and the $\mathcal{K}_2(x)$-kink is stable. 
        \end{itemize}

As a conclusion of the analysis presented in this section, we may state that the model under  consideration supports two types of basic kinks. In the regime $\mu\in (0,1)$, the kink $\mathcal{K}_{AB}(x)$ defined in the topological sector $AB$ consists of a single energy lump and can therefore be regarded as one of the fundamental extended particles of the theory. Likewise, in this same regime, the basic kink $\mathcal{K}_1(x)$ in the topological sector $AA$ is also formed by a single lump and constitutes the second fundamental extended particle of the model. In addition to these solutions, there exists a one-parameter family of kinks in the sector $BB$, namely the $\mathcal{K}_{B\overline{B}}(x,\kappa)$-solutions. For $\mu\in (0,1)$, this family includes the configuration with only one non-vanishing field component, denoted by $\mathcal{K}_2(x)$. The $\mathcal{K}_{B\overline{B}}(x,\kappa)$-family can be interpreted as a nonlinear superposition of three basic extended particles: two $\mathcal{K}_{AB}(x)$-kinks symmetrically located at equal distances from a central $\mathcal{K}_1(x)$-kink. An interesting feature arises from the stability properties of the solution $\mathcal{K}_1(x)$. For $\mu>\frac{1}{4}$, this kink becomes unstable. In this case, one would expect it to decay into two $\mathcal{K}_{AB}(x)$-type lumps, so that the composite configuration described above would ultimately consist of four $\mathcal{K}_{AB}(x)$-lumps. In contrast, when $\mu<\frac{1}{4}$, the kink $\mathcal{K}_1(x)$ is stable. Under these conditions, intersolitonic forces between the two $\mathcal{K}_{AB}(x)$-kinks in the sector $AA$ may reduce their separation, leading to a configuration in two of these kinks evolves into a $\mathcal{K}_1(x)$-type lump. The borderline case $\mu=\frac{1}{4}$ corresponds to a situation of neutral equilibrium: the superposition of lumps can be maintained at arbitrary separation, provided that the two $\mathcal{K}_{AB}(x)$-lumps remain symmetrically placed with respect to the central $\mathcal{K}_1(x)$-kink. 

An entirely analogous analysis, exploiting the symmetry between the regimes $\mu \in (0,1)$ and $\mu>1$, can be carried out for the complementary parameter range, leading to a qualitatively equivalent scenario.

\end{itemize}

\section{Models admitting superpotentials with two singular points}

Finally, we explore the possibility of identifying models with fourth-order polynomial potentials derived from superpotentials that involve an irrational function
\begin{equation}
    W(\phi_1,\phi_2) = \sqrt{((\phi_1-a_1)^2+(\phi_2-a_2)^2)((\phi_1-a_3)^2+(\phi_2-a_4)^2)}\,\,(b_1\,\phi_1 + b_2\,\phi_2 + b_3),\,\,\,\, a_i,b_j\in\mathbb{R}\,.
    \label{eq:2singpts}
\end{equation}
 with two singular isolated points, which are located at the points $(a_1,a_2)$ y $(a_3,a_4)$. Following the same procedure as in section 4, we can identify two distinct systems, which we list below:
\begin{itemize}
    \item The first of these possibilities fixes the coefficients $a_i$ and $b_i$ such that the resulting superpotential $W$ is given by the expression 
\begin{equation}
    W_5 (\phi_1,\phi_2)= \phi_2\,\sqrt{\phi_1^4 + 2\,\phi_1^2\,(\phi_2^2 - 1) + (1 + \phi_2^2)^2};
\end{equation}
     which ultimately yields the single potential term written as:
    \begin{equation}
        U_5(\phi_1,\phi_2) = \frac{1}{2}(1-\phi_1^2)^2 + \frac{1}{2}(1 + 3\,\phi_2^2)^2 + 5\,\phi_1^2\,\phi_2^2-\frac{1}{2},
    \end{equation}

\item The second possibility is characterized by the superpotential 
\begin{equation}
    W_6(\phi_1,\phi_2) = \frac{\phi_1}{3}\,\sqrt{\phi_1^4 + 2\,\phi_1^2\,(\phi_2^2 - 3) + (3 + \phi_2^2)^2}
\end{equation}
which defines the expression for the potential V as follows:
\begin{equation}
    U_6 (\phi_1,\phi_2)= \frac{1}{2}(1-\phi_1^2)^2 + \frac{1}{18}(3 + \phi_2^2)^2 + \frac{10}{18}\phi_1^2\,\phi_2^2-\frac{1}{2}\,.
\end{equation}
\end{itemize}
Note that these systems do not conform families of models; the expressions only involve numerical coefficients. Besides, both models have only two vacua $A_\pm = (\pm1,0)$ and, in addition, these are the points at which $W_5$ and $W_6$ become singular. As a result, the orbits solutions to the gradient equations associated with these superpotentials are never closed, and therefore, they do not include any other topological kink different from the pre-established one non-null component kink (\ref{TK1}).

\section{Confluences between models with different superpotentials}

Under certain circumstances, a single model may admit two or more superpotentials, a situation previously noted in Wess-Zumino-type models. When this occurs, the Bogomolny arrangement can be achieved in different ways, each leading to a set of first-order differential equations that are, in principle, distinct. This, in turn, implies the possible coexistence of two or more one-parameter families of kink solutions. The purpose of this section is to investigate whether models exist that possess several superpotentials among those analyzed in previous sections; in other words, we aim to determine if different superpotentials among those studied lead to the same potential. In this regard, we highlight the following cases where multiple kink families coexist simultaneously within the same model:

\begin{itemize}
    \item \textit{The BNRT model with $\beta=1$:} If we consider the potential (\ref{eq:POTP2}) for the particular coupling constant $\beta=1$, that is,
    \begin{equation}
     U(\phi_1,\phi_2)=\frac{1}{2}(1-\phi_1^2)^2+\frac{1}{2}(1- \phi_2^2)^2+ 3 \,\phi_1^2\,\phi_2^2-\frac{1}{2}  \label{conflu1}
    \end{equation}
    we observe that it can be derived from two distinct superpotentials, namely from the superpotential (\ref{eq:SUPP2}) with $\beta=1$, $W_2(\phi_1,\phi_2)=\frac{1}{3} \phi _1 \left(3 \, \phi _2^2+\phi _1^2-3\right)$ and (\ref{eq:SUPP3}) with $\gamma=1$, $\widetilde{W}(\phi_1,\phi_2) = \frac{1}{3} \phi_2 (-\phi_2^2- 3 \phi_1^2 +3)$. The potential (\ref{conflu1}) possesses a vacuum manifold consisting of four isolated minima located at ${\cal M}=\{A_\pm =(\pm 1, 0), \, B_\pm =(0,\pm 1)\}$. As shown in Section 3, the superpotential (\ref{eq:SUPP2}) gives rise to a one-parameter family of kink solutions in the topological sector $AA$. These solutions can be written explicitly as 
    \begin{equation}
        \mathcal{K}_{A\overline{A}}^{(\alpha,\beta)}(x;b)=\left((-1)^{\alpha}\frac{\sinh2\bar{x}}{b+\cosh\bar{x}},(-1)^{\beta}\frac{\sqrt{b^2-1}}{b+\cosh\bar{x}}\right)\,,
    \end{equation}
    and interpolate between the corresponding pair of vacua within that sector. Here, $b\in(1,\infty)$. In an analogous manner, the first-order equations associated with the alternative superpotential (\ref{eq:SUPP3}) can also be solved explicitly. This construction yields a new family of kink solutions that connect the vacua $B_\pm$ and $B_\mp$, which are expressed as  
    \begin{equation}
        \mathcal{K}_{B\overline{B}}^{(\alpha,\beta)}(x;b)=\left((-1)^{\alpha}\frac{\sqrt{b^2-1}}{b+\cosh\bar{x}},(-1)^{\beta}\frac{\sinh2\bar{x}}{b+\cosh\bar{x}}\right)\,.
    \end{equation}
    Notice that $\mathcal{K}_1(x)=\mathcal{K}_{A\overline{A}}(x;1)$ whereas $\mathcal{K}_2(x)=\mathcal{K}_{B\overline{B}}(x;1)$. Moreover, we can obtain the kinks in the $AB$ sector taking the limit $b\to \infty$ in whichever $\mathcal{K}_{A\overline{A}}(x;b)$ or $\mathcal{K}_{B\overline{B}}(x;b)$, indeed
    \begin{equation}
        \mathcal{K}_{AB}^{(\alpha,\beta)}(x)=\left((-1)^{\alpha}(1-\tanh\bar{x}),(-1)^{\beta}(1+\tanh\bar{x})\right).
    \end{equation}
    Hence, now the family parameter $b$ determines the separation between the elementary particles of the model. The energy sum rule clearly reveals the two-lumps structure in this case,
    \begin{equation}
       E[\mathcal{K}_{A\overline{B}}(x)]=\frac{2}{3}\qquad ,\qquad E[\mathcal{K}_{A\overline{A}}(x;b)]=E[\mathcal{K}_{B\overline{B}}(x;b)]=2E[\mathcal{K}_{A\overline{B}}(x)].
    \end{equation}

    \item \textit{The BNRT model with $\beta=\frac{1}{4}$:} It is straightforward to check that the potential $U_2$ and $U_4$ coincides when $\beta=\mu=1/4$,
    \begin{equation}
        U(\phi_1,\phi_2)=\frac{1}{2}(1-\phi_1^2)^2+\frac{1}{32}\left(4-\phi_2^2\right)^2+\frac{3}{8}\,\phi_1^2\,\phi_2^2-\frac{1}{2}
        \label{eq:conflu2}
     \end{equation}
which means that \eqref{eq:conflu2} can be derived from the following two superpotencials
\begin{align}
    & W_I(\phi_1,\phi_2) = \frac{1}{3}\sqrt{\phi_1^2+\phi_2^2}\,\Big(\phi_1^2+\frac{\phi_2^2}{4}-3\Big), \label{WI}\\
    & W_{II}(\phi_1,\phi_2) =\frac{1}{3} \phi _1 \left(\frac{3}{4}\, \phi _2^2+\phi _1^2-3\right). \label{WII}
\end{align}
In this case, the vacuum manifold include four elements 
\[
{\cal M} = \left\{ A_\pm = (\pm 1, 0) \hspace{0.2cm}, \hspace{0.4cm} B_\pm =(0, \pm \, 2) \right\}
\]
From the previous results, it can be concluded that a one-parameter family of kinks exists in the $AA$ -topological sector, asymptotically connecting points $A_\pm$ and $A_\mp$, alongside another one-parameter family in the $BB$ topological sector connecting points $B_\pm$ and $B_\mp$. In this particular case, the spatial dependence of the solutions can be explicitly calculated. Thus, the first family is expressed as 
\begin{equation}
    \mathcal{K}^{(\alpha,\beta)}_{AA}(x;b)= \Big( (-1)^{\alpha}\frac{\sinh \bar{x}}{b^2+\cosh \bar{x}} \hspace{0.1cm},\hspace{0.1cm} (-1)^{\beta}\frac{2 b}{\sqrt{b^2+\cosh \bar{x}}} \Big), \label{eq:KAA}
\end{equation}
where the family parameter $b\in (1,\infty)$. On the other hand, the second is given by 
\begin{equation}
    \mathcal{K}^{(\alpha)}_{BB}(x;c) = \Big( \frac{(-1)^{\alpha} 2\sinh 2 \sqrt{2}\, c\,\sinh \bar{x}}{2 \cosh 2 \sqrt{2} \,c\, \cosh \bar{x}+\cosh ^2\bar{x}+1},\frac{ (-1)^{\alpha+1} 2 \sinh \bar{x}}{\sqrt{2 \cosh 2 \sqrt{2}\, c\,\cosh \bar{x}+\cosh ^2\bar{x}+1}} \Big) \label{eq:KBB}
\end{equation}
with $c\in \mathbb{R}$. As a matter of fact, the relation between the family parameters $c$ and $\kappa$ is $\kappa = 8\Big(1-\text{coth}\,2\sqrt{2}\,c\Big)$. Hence, for $\kappa\to\kappa_c(1/4)=0$, $c\to\infty$, which corresponds to the case where the three lumps are infinitely separated. Conversely, when $c\to0$, $\kappa\to\infty$. In addition to these families, eight isolated topological kinks/antikinks can be found in the $AB$ sector, whose expressions are provided below: 
\begin{equation}
    \mathcal{K}^{(\alpha,\beta)}_{AB}(x) = \left( (-1)^{\alpha}\frac{1}{2}\left(1-\tanh\frac{\bar{x}}{2}\right),(-1)^{\beta}\sqrt{2\Big(1+\tanh\frac{\bar{x}}{2}\Big)} \right)
\end{equation}
Furthermore, the calculation of the total energies for these solutions is highly illustrative: 
\[
E[\mathcal{K}_{AB}(x)] = \frac{2}{3},\qquad E[\mathcal{K}_{A\overline{A}}(x;b)] = 2\,E[\mathcal{K}_{AB}(x)]=\frac{4}{3}; 
\]
\[
E[\mathcal{K}_{B\overline{B}}(x;c)] = 2\,E[\mathcal{K}_{A\overline{A}}(x)]=4\,E[\mathcal{K}_{AB}(x)]\,.
\]
It can be observed that all members $\mathcal{K}_{A\overline{A}}(x;b)$ of the $AA$ family share the same energy, which is twice that of the kinks $\mathcal{K}_{AB}(x)$, whereas the kinks $\mathcal{K}_{B\overline{B}}(x;c)$ in the $BB$ sector possess four times that energy. The energy distribution indicates that the $\mathcal{K}_{AB}(x)$-kinks exhibit a single energy lump, suggesting that these solutions describe the basic extended particles of the model. Conversely, the $\mathcal{K}_{A\overline{A}}(x;b)$-kinks can be interpreted as two basic particles separated by a distance $d$ determined by the parameter value (which may even overlap). Finally, the $\mathcal{K}_{B\overline{B}}(x;c)$-kinks can be seen as a configuration of four such basic particles: two separated by the same distance from a central core, which itself results from the overlap of two basic particles.

It should be noted that these interpretations had already been discussed separately for the models associated with the superpotentials (\ref{WI}) and (\ref{WII}). However, in the present case this double interpretation can be implemented within a single model. This implies that two fundamental extended particles of type $\mathcal{K}_{AB}(x)$ may be placed at arbitrary separation without experiencing intersolitonic forces. Moreover, an analogous situation arises when two $\mathcal{K}_{AB}(x)$-type extended particles are located symmetrically at equal distances from a central lump of type $\mathcal{K}_1(x)$. The distinctive feature in comparison with cases 2 and 4 analyzed previously is that the $\mathcal{K}_1(x)$-lump can no longer be regarded as a fundamental extended particle. Indeed, it can be interpreted instead as a composite configuration formed by two overlapping $\mathcal{K}_{AB}(x)$ lumps.
\end{itemize}

We would like to emphasize that the confluence cases identified here confirm the hypothesis formulated in Section 3, where we anticipated the possible existence of a second family of kinks for particular values of the coupling constant $\beta$ in the model (\ref{eq:POTP2}), namely $\beta=1,\frac{1}{4},\frac{1}{8},\frac{1}{13}, \dots$. In the present section, we have explicitly demonstrated that this expectation is fulfilled indeed for the cases $\beta=1$ and $\beta = \frac{1}{4}$. The explicit construction of the corresponding kink families for the remaining values of $\beta$ remains an open problem. It is reasonable to expect such solutions to arise from superpotentials with a more intricate functional structure than those considered in this work.
  \begin{figure}[h!]
  \centering
    \includegraphics[width=6cm]{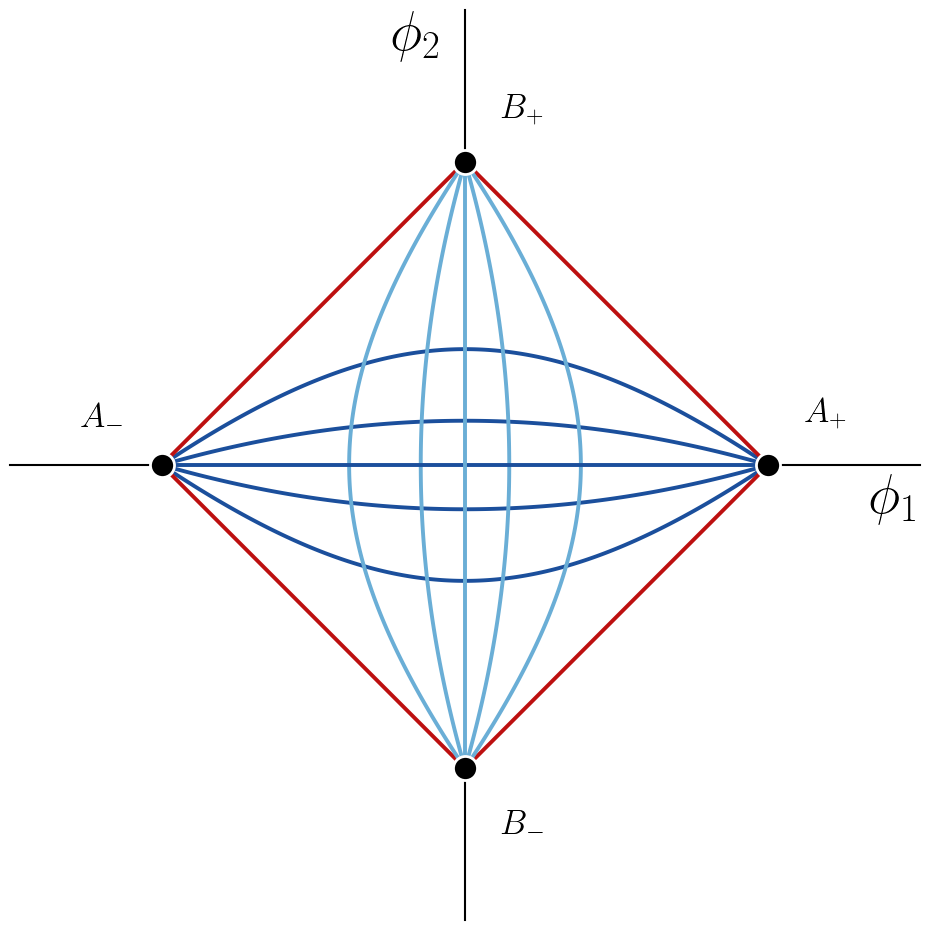} \hspace{1cm}
    \includegraphics[width=6cm]{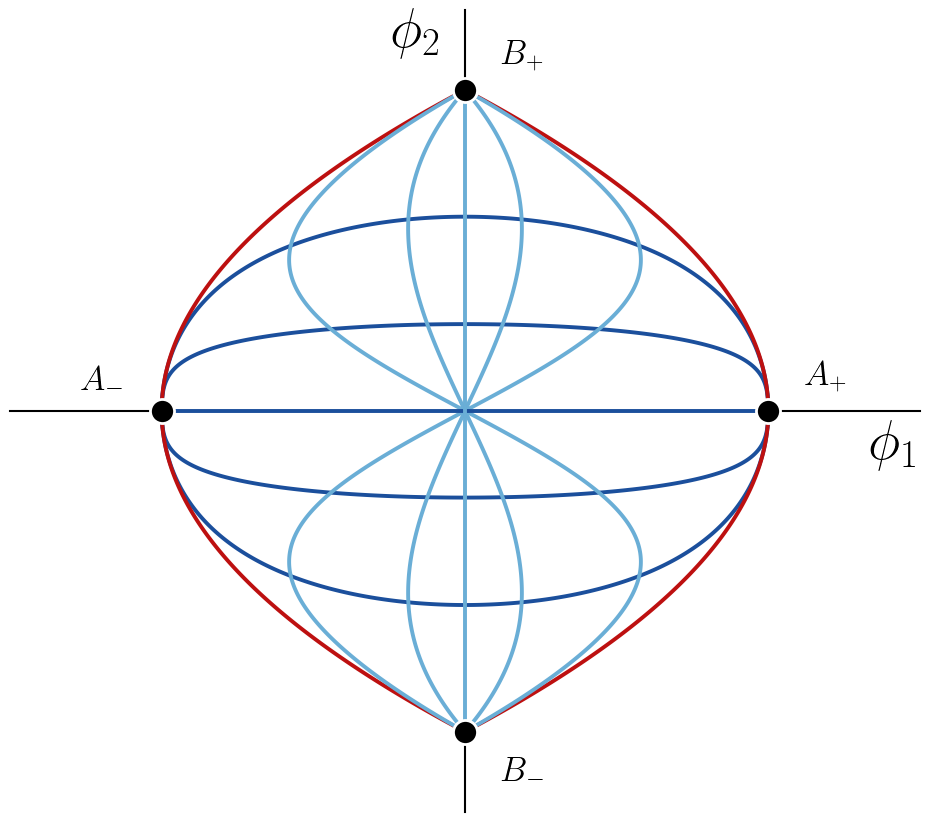}
  \caption{Kink orbits in the $AA$ sector (dark blue), $BB$ sector (light blue) and $AB$ sector (red) for the special cases BNRT $\beta=1$ (left) and BNRT $\beta=1/4$ (right). The black dots represent the vacua in each case.}
  \label{fig:U5orb}
\end{figure}

\section{Conclusions}

In this work, we have carried out a systematic and constructive analysis of two-component scalar field theories in (1+1) dimensions with self-interaction terms of at most quartic order and endowed with a $\mathbb{Z}_2\times \mathbb{Z}_2$-symmetry which admit continuous families of BPS and semi-BPS kinks. Our main objective has been to clarify whether the examples previously known in the literature exhaust all possible realizations within this framework. While cubic polynomial superpotentials naturally generate quartic scalar interactions, we have shown that more general, including irrational, functional forms may also lead to admissible quartic models. This observation considerably enlarges the landscape of analytically tractable theories and reveals new models unexplored in the literature.

Within this unified setting, the classical MSTB and BNRT models emerge as particular representatives in this procedure. By revisiting these benchmark theories, we have placed them in a common structural framework and clarified how their kink families arise from specific superpotential choices. More importantly, our constructive method has led to genuinely new models. In particular, the case derived from an irrational superpotential with a single singular point exhibits semi-BPS kink families whose structure differs qualitatively from that of the standard polynomial models. The presence of a non-differentiable point in the superpotential allows for solutions that satisfy different first-order equations in distinct regions of internal space, giving rise to novel moduli spaces of defects. In particular, the members of kink families can be interpreted as composite objects. The solutions belonging to a continuous family are most naturally understood as nonlinear superpositions of fundamental one-lump kinks, whose relative separation is controlled by the modulus parameter. The associated energy sum rules provide a precise quantitative formulation of this composite picture. Depending on the regime of the coupling constants, the role of the basic constituents may change: a given one-component kink can be either a fundamental stable particle or an unstable configuration that decomposes into more elementary lumps. In this way, the internal structure and physical interpretation of the defects evolve continuously across parameter space.

Another significant result is the identification and analysis of the phenomenon that we have termed \textit{confluence}, whereby a single scalar potential can be generated by two or more non-trivially related superpotentials. This property implies that one and the same model may host several distinct kink families, thereby enriching the stationary solution manifold beyond what is possible in theories derived from a unique superpotential. Special parameter values correspond to neutral equilibrium configurations in which constituent lumps may be placed at arbitrary separation without experiencing intersolitonic forces, signaling an enhanced degeneracy of the classical configuration space.

The stability analysis of the various solutions further clarifies the interplay between coupling constants and defect structure. Continuous kink families are characterized by the presence of two zero modes in the fluctuation spectrum: the translational mode and an additional mode associated with the family modulus. Changes in stability at critical parameter values determine whether certain one-component kinks act as stable fundamental constituents or as unstable configurations that decay into composite states. These results provide a coherent dynamical interpretation of the different regimes identified in our classification.

Overall, our work enlarges the catalogue of two-component quartic scalar field theories with analytically accessible kink families and establishes a systematic framework for their construction and interpretation. Several extensions of the present work can be envisaged. First, an open problem concerns the possible existence of kink families in the $BB$ topological sector of the BNRT model for the particular values of the coupling constant $\beta=\frac{1}{8},\frac{1}{13},\dots$, where the presence of two zero modes was identified and once time that this has been proved for the first two cases $\beta=1,\frac{1}{4}$ given in (\ref{casesBNRT}). Beyond this specific issue, the procedure developed here can be naturally generalized to broader settings. In particular, one may consider models with a larger number of scalar field components, which are expected to support even richer vacuum structures and moduli spaces. Likewise, the restriction to quartic interactions may be relaxed by allowing polynomial potentials of sixth degree. Such an extension would likely generate a substantially wider class of models, potentially revealing new families of topological defects that have not yet been explored in the literature.

\section*{Acknowledgments}

The authors have been supported in part by Spanish Ministerio de Ciencia e Innovación (MCIN) with funding from the grant PID2023-148409NB-I00 MTM, and they were also partially supported by Fundaci\'on Sol\'orzano through the project FS/11-2024.

\section*{Conflict of Interest}
The authors declare that they have no conflict of interest.
 
\section*{Data availibility}
Data sharing is not applicable to this article as no datasets were generated or analysed during
the current study.


\end{document}